\documentclass[a4paper,11pt]{article}
\pdfoutput=1
\usepackage{jheppub} 
\usepackage{axodraw2}
\usepackage{amsmath}
\usepackage{breqn}
\usepackage{url}

\usepackage{pgfplots}
   \pgfplotsset{compat=1.5}
   \usepgfplotslibrary{fillbetween}

\definecolor{dgreen}{HTML}{008000}

\definecolor{dblue}{HTML}{0000A0}

\title{Hadron Collider Sensitivity to Fat Flavourful $Z^\prime$s for $R_{K^{(\ast)}}$}  
\author[a]{B.C. Allanach,}
\author[b]{Tyler Corbett,\footnote{Corresponding author.}}
\author[b]{Matthew J. Dolan,}
\author[a,c]{Tevong You,}

\affiliation[a]{DAMTP, University of Cambridge, Wilberforce Road, Cambridge, 
CB3 0WA, United Kingdom}
\affiliation[b]{School of Physics, University of Melbourne, Victoria 3010,
  Australia} 
\affiliation[c]{Cavendish Laboratory, University of Cambridge, J.J. Thomson Avenue, Cambridge, CB3 0HE, United Kingdom}

\emailAdd{B.C.Allanach@damtp.cam.ac.uk}
\emailAdd{corbett.t.s@gmail.com}
\emailAdd{dolan@unimelb.edu.au}
\emailAdd{tty20@cam.ac.uk}
\preprint{Cavendish-HEP-2018-14,  DAMTP-2018-33}
\abstract{We further investigate the case where  new physics in the form of a
  massive   $Z^\prime$ particle explains apparent measurements 
  of lepton flavour non-universality in $B \rightarrow K^{(\ast)} l^+ l^-$
  decays. Hadron collider sensitivities for direct production of such
  $Z^\prime$s  have been previously studied
  in the narrow width 
  limit for a $\mu^+ \mu^-$ final state. Here, we extend the analysis
  to sizeable decay widths and improve the
  sensitivity 
  estimate for  the narrow width case.
  We estimate the sensitivities of
  the high luminosity 14 TeV Large 
  Hadron Collider (HL-LHC), a high energy 27 TeV LHC (HE-LHC), as well as a
  potential 100 TeV future circular collider (FCC). The HL-LHC has sensitivity
  to narrow $Z^\prime$ resonances consistent with the anomalies. In one of our
  simplified models the 
  FCC could probe 23 TeV $Z^\prime$ particles with widths of up to 
  0.35 of their mass at 95\% confidence level (CL). In another model, the
  HL-LHC and HE-LHC cover sizeable portions of parameter space, but the
  whole of perturbative parameter space can be covered by the FCC. 
}

\begin{document}
\maketitle 
\flushbottom

\section{Introduction}
\label{sec:introduction}

Over the past number of years, there has been much interest in a number of
 anomalies\footnote{In the present paper, we use `anomaly' to refer to a
   tension between an experimental measurement and its Standard Model
   prediction.} in flavour physics. Specifically, the ratio of branching
 ratios (BRs)
\begin{equation}
R_K \equiv
 \frac{BR(B\rightarrow K \mu^+\mu^-)}{BR(B\rightarrow
  K e^+e^-)}, \qquad
R_{K^\ast} \equiv \frac{BR(B\rightarrow K^{\ast}\mu^+\mu^-)}{BR(B\rightarrow
  K^{\ast}e^+e^-)}. 
\end{equation}
have substantial deviations from
 Standard Model (SM) expectations~\cite{Aaij:2014ora,Aaij:2017vbb}. 
There are also discrepancies in the angular
 variable $P_5'$~\cite{Aaij:2013qta,Aaij:2015oid} and $BR(B_s \rightarrow \phi \mu^+\mu^-)$. It is by now well known
 that it is possible to account for these anomalies through the existence of
 new physics which contributes to the neutral current $b\to s \mu^+\mu^-$ decay
 channel~\cite{Altmannshofer:2013foa,Altmannshofer:2014rta,Altmannshofer:2017yso,Ciuchini:2017mik,Capdevila:2017bsm,Geng:2017svp,DAmico:2017mtc,DiChiara:2017cjq}. 

The most well-studied UV-complete explanations of these anomalies involve
either flavour-violating $Z^\prime$s and/or leptoquarks. We shall focus here on the
$Z^\prime$ scenario. Models falling into this category involve a new gauge
group beyond 
the SM\@. This could be an abelian extension such as
$L_\mu-L_\tau$ and related gauge
groups~\cite{Gauld:2013qba,Buras:2013dea,Buras:2013qja,Altmannshofer:2014cfa,Buras:2014yna,Crivellin:2015mga,Crivellin:2015lwa,Sierra:2015fma,Crivellin:2015era,Celis:2015ara,Greljo:2015mma,Altmannshofer:2015mqa,Allanach:2015gkd,Falkowski:2015zwa,Chiang:2016qov,Becirevic:2016zri,Boucenna:2016wpr,Boucenna:2016qad,Ko:2017lzd,Alonso:2017bff,Alonso:2017uky,1674-1137-42-3-033104, Ellis:2017nrp, CHEN2018420,Faisel:2017glo,PhysRevD.97.115003,Bian:2017xzg,PhysRevD.97.075035,Duan:2018akc},
or the new gauge group could be non-abelian~\cite{Buras:2013dea}, leading to
the existence of $W^\prime$ particles, for example. There are also models with multiple
abelian groups~\cite{Crivellin:2016ejn} leading to multiple $Z^\prime$ particles. Most of these models involve generating the $b\to s\mu^+\mu^-$ transition at tree-level, although a loop-level penguin is also possible~\cite{Kamenik:2017tnu}, which requires a much lighter $Z^\prime$ due to the loop suppression.

While one can study the effects of these models on flavour physics indirectly using effective field theory, one would also like to pin down the properties
of the new resonances through their direct production in a high energy
collider environment. This raises the exciting prospect of 
\begin{quote}{\em directly experimentally probing the new physics that explains aspects of the fermion mass
  problem}~\cite{King:2018fcg,Allanach:2018lvl,Grinstein:2018fgb} \end{quote} 
(i.e.\ the
patterns and hierarchies in fermion masses and 
mixing parameters). It has previously been argued that perturbative
unitarity requires that the new physics responsible for the flavour anomalies
must enter at a scale below 80~TeV~\cite{Altmannshofer:2017yso}. Other
phenomenological bounds, notably from the measurement of $B_s -
\overline{B_s}$ mixing, imply a stricter upper bound for perturbative values
of the $Z^\prime$ coupling if one wants to simultaneously fit $R_{K^{(\ast)}}$.
Accordingly,
one may hope that the resonances may be accessible at a future hadron
collider, or at the High-Luminosity Large Hadron Collider (HL-LHC). 

In Ref.~\cite{Allanach:2017bta}, sensitivities of future hadron colliders were
estimated for particles that can explain the
$R_{K}$ and $R_{K^{(\ast)}}$ measurements. 
In particular, 
 Ref.~\cite{Allanach:2017bta} considers
 the case where the anomalies are explained by a $Z^\prime$
 or leptoquark, each with flavour dependent couplings. The $\mu^+ \mu^-$
 channel was used for the $Z^\prime$ case.
 Simplifications in the
 analysis (an extrapolation of current 
 LHC search limits, assuming that acceptances and efficiencies don't
 change with centre of mass energy) 
 required also that the decay width of a new $s-$channel
 resonance was narrow
 (defined to be less than 10\% of its mass). However, a substantial region of
 the parameter 
 space which fits the $B-$anomalies requires large $\mathcal{O}(1)$
 couplings which lead to large decay widths. 
 Moreover, the high-luminosity run of the LHC may yet see indirect signs of new
 physics from effective operators in the high invariant mass tail of di-lepton
 distributions~\cite{Greljo:2017vvb,Alioli:2017nzr}. We shall see that such a
 large effect would typically imply an underlying wide resonance. 
 
 Therefore, in this paper, we study
 the reach and implications of the fat, flavourful $Z^\prime$ scenario, taking
  effects of the large width into account in our simulations. By simulating
 $Z^\prime$ signal events,
 we also take 
 into account changes in acceptances and efficiencies when operating at
 different centre of mass energies. 
 We study the phenomenology
 of two $SU(2)_L$ invariant simplified models, which we dub the Mixed Up-Muon
 and Mixed Down-Muon models, leaving the study of large-width leptoquarks
 to future work\footnote{We note that other types of new particles have been
   proposed to resolve the 
   tension between 
measurements and SM predictions of $B$ to $D^{(\ast)} \tau \nu$
decays~\cite{Lees:2012xj,Lees:2013uzd,Aaij:2015yra,Huschle:2015rga,Sato:2016svk,Hirose:2016wfn,Hirose:2017dxl}. These
particles (e.g.\ $W^\prime$s or other types of leptoquark)
must be much lighter or much more strongly coupled than the ones responsible
for the $bs\mu^+\mu^-$ 
anomalies in order to fit data, and so should be consequently
easier to detect. The study of these other types of particles is also left to
future work.}. 

We focus our attention on the HL-LHC,
the High-Energy LHC (HE-LHC), and the Future Circular Collider (FCC)\@. We find that while the HL-LHC is sensitive only to
narrow ($\Gamma_{Z^\prime}/M_{Z^\prime}<0.1$) resonances, the HE-LHC and FCC
could probe fat, flavourful resonances with widths of up to 35\% of the mass
for $Z^\prime$ masses up to 23~TeV for the Mixed-Up Muon model, and the entire
perturbative region of parameter space for the Mixed-Down Muon model.

We proceed as follows:
in \S\ref{sec:simp} we develop simplified $Z^\prime$ models for the
$B-$anomalies and 
detail the other important
constraints such as meson mixing, neutrino trident
production, and indirect effects in the di-muon invariant mass distribution. In \S\ref{sec:sens} we present our projections on future collider
sensitivity to these models before concluding in \S\ref{sec:conc}. 
Our notation for the fields is listed in Appendix~\ref{sec:def} by
detailing their SM quantum numbers.

\begin{figure}
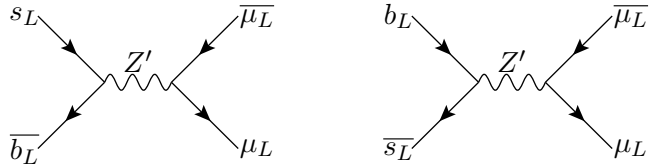

\begin{center}
\begin{axopicture}(200,50)(-40,0)
\Line[arrow](100,50)(125,25)
\Line[arrow](125,25)(100,0)
\Line[arrow](175,50)(150,25)
\Line[arrow](150,25)(175,0)
\Photon(125,25)(150,25){3}{3}
\Text(137.5,33)[c]{$Z^\prime$}
\Text(95,50)[c]{$b_L$}
\Text(95,0)[c]{$\overline{s_L}$}
\Text(182,50)[c]{$\overline{\mu_L}$}
\Text(182,0)[c]{$\mu_L$}
\Line[arrow](-40,50)(-15,25)
\Line[arrow](-15,25)(-40,0)
\Line[arrow](35,50)(10,25)
\Line[arrow](10,25)(35,0)
\Photon(-15,25)(10,25){3}{3}
\Text(-2.5,33)[c]{$Z^\prime$}
\Text(-45,50)[c]{$s_L$}
\Text(-45,0)[c]{$\overline{b_L}$}
\Text(42,50)[c]{$\overline{\mu_L}$}
\Text(42,0)[c]{$\mu_L$}
\end{axopicture}
\end{center}
\caption{Feynman diagrams of parton interactions in $pp$ collisions where
  a flavourful $Z^\prime$ produces a $\mu^+\mu^-$ final state. In the low
  momentum limit, the same diagrams 
  generate an
  effective operator capable of accounting for the discrepancies in $B \to
  K^{(*)} \mu^+ \mu^-$ decays as compared to SM predictions. \label{fig:feyn}} 
\end{figure}

\section{Simplified Models }
\label{sec:simp}

We consider two representative models of $Z^\prime$s,
following Ref.~\cite{Allanach:2017bta}, which introduced the na\"ive and the $33\mu\mu$
models. The tree-level $Z^\prime$ Lagrangian couplings that we know must be present
in $Z^\prime$ models 
in order to explain the neutral current $B-$anomalies are
\begin{equation}
\mathcal{L}_{Z^\prime f} =\left( g_{sb} Z^\prime_{\rho} \overline{s_L} \gamma^{\rho} b_L  +
  \text{h.c.} \right)  + g_{\mu\mu} Z_{\rho}' \overline{\mu_L}\gamma^\rho \mu_L + 
\ldots \label{wrongNaive}
  \end{equation}
A fit to $R_{K^{(\ast)}}$ and other `clean\footnote{i.e.\ observables with small
theoretical uncertainties in their predictions.}' $B-$anomalies in
Ref.~\cite{DAmico:2017mtc} found that   
the couplings and masses of $Z^\prime$ particles are constrained to be
\begin{equation}
g_{bs}g_{\mu\mu} = -x \left(\frac{M_{Z^\prime}}{31\text{TeV}}
\right)^2,
\label{constraint}
  \end{equation}
if $g_{bs}$ and $g_{\mu\mu}$ are real, where $x=1.00 \pm 0.25$. 
Throughout this paper, we shall enforce Eq.~\ref{constraint}, taking the
central value $x=1.00$ from the fit. 
In general, $g_{bs}$ and $g_{\mu\mu}$ are complex. However, here, we take
$g_{\mu\mu}$ to be real and positive and $g_{bs}$ to be negative.
In the models we introduce below, $g_{bs}$
may have a 
small imaginary part. Since the full effects of complex phases
are outside the scope of this work, whenever we refer to $g_{bs}$ below, we
shall implicitly refer to its absolute value.

\subsection{Constraints}

\begin{figure}
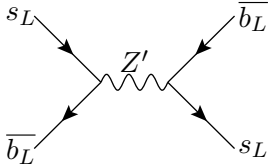

\begin{center}
\begin{axopicture}(50,50)(-40,0)
\Line[arrow](-40,50)(-15,25)
\Line[arrow](-15,25)(-40,0)
\Line[arrow](35,50)(10,25)
\Line[arrow](10,25)(35,0)
\Photon(-15,25)(10,25){3}{3}
\Text(-2.5,33)[c]{$Z^\prime$}
\Text(-45,50)[c]{$s_L$}
\Text(-45,0)[c]{$\overline{b_L}$}
\Text(42,50)[c]{$\overline{b_L}$}
\Text(42,0)[c]{$s_L$}
\end{axopicture}
\end{center}
\caption{Feynman diagram of the tree-level $Z^\prime$ contribution to $B_s-\overline{B_s}$
  mixing. \label{fig:dms}}  
\end{figure}
$Z^\prime$ models are subject to a number of constraints,
the strongest being from measurements of  $B_s-\overline{B_s}$ mixing, which
constrains a function of $g_{bs}$ and $M_{Z^\prime}$.
A Feynman diagram depicting the $Z^\prime$ contribution is shown in
Fig.~\ref{fig:dms}. 
Another constraint comes from 
neutrino trident
production which is sensitive to $g_{\mu\mu}$ and $M_{Z^\prime}$. We adapt the
bound on 
$B_s$-mixing from Ref.~\cite{Arnan:2016cpy}, using the $2\sigma$ constraint derived
from the 2016 FLAG average on the hadronic form factor $f_{B_s}$ and bag
parameter $B_{B_s}$. More recently, the
Fermilab/MILC Collaboration has presented a new
determination~\cite{Bazavov:2016nty} of these non-perturbative parameters
substantially higher than 
previous results which means that the 
$B_s-\overline{B_s}$ mixing measurement would be in tension with inferred SM
predictions. This 
would imply a much 
stronger bound on $Z^\prime$s (which would have the wrong sign contribution to
explain the 
tension)~\cite{DiLuzio:2017fdq}. However, Ref.~\cite{Kumar:2018kmr} has
observed that the Fermilab/MILC result is large enough to have important
implications for $\Delta M_d$ and the CKM matrix. Accordingly, we take the
constraint of Ref.~\cite{Arnan:2016cpy} as our primary one, but also show the
constraint of Ref.~\cite{DiLuzio:2017fdq}. We find that the
result from Ref.~\cite{DAmico:2017mtc} is equivalent to\footnote{In the
  analysis of Ref.~\cite{DiLuzio:2017fdq}, the equivalent bound is much
  stronger: $|g_{bs}|\lesssim M_{Z^\prime}/(600\text{~TeV})$.}  
\begin{equation}
|g_{bs}|\lesssim
M_{Z^\prime}/(148\text{~TeV}), \label{bsmix}
\end{equation} which is the bound that we use here.
Eqs.~\ref{constraint} and~\ref{bsmix} imply
\begin{equation}
\frac{g_{\mu\mu}}{|g_{bs}|} \gtrsim 31x. \label{relstr}
\end{equation}

We also take into account neutrino trident production, $\nu_{\mu} \gamma^{*}
\to \nu_\mu \mu^+ \mu^-$, using the constraints on leptonic operators in the
SM effective field theory
(EFT) from Ref.~\cite{Falkowski:2017pss,Falkowski:2018dsl}. This corresponds to
$g_{\mu\mu} \lesssim {M_{Z^\prime}}/({0.39\text{~TeV}})$ which
sets an upper bound on the muon coupling for low $m_{Z^\prime}$. This upper
bound is not
strong enough to 
affect our projections to future colliders. 

\subsection{Model definition and couplings}
Including only the $Z^\prime$ couplings in Eq.~\ref{wrongNaive} without the ellipsis
was called the
na\"ive model in 
Ref.~\cite{Allanach:2017bta}. However, we now introduce a new similar
simplified model
that respects $SU(2)_L$. In order to do this, we must first set up 
the mass eigenbasis and the weak eigenbasis. Writing the SM weak
eigenbasis 
fermionic fields with a prime (see appendix~\ref{sec:def}
for the field definitions): 
$$
{\bf u_J'}=\left( \begin{array}{c} u_J' \\ c_J' \\ t_J' \\ \end{array}
\right), \qquad
{\bf d_J'}=\left( \begin{array}{c} d_J' \\ s_J' \\ b_J' \\ \end{array} \right), \qquad
{\bf n_L'}=\left( \begin{array}{c} {\nu_e}_L' \\ {\nu_\mu}_L' \\ {\nu_\tau}_L' \\ \end{array} \right), \qquad
{\bf e_J'}=\left( \begin{array}{c} e_J' \\ \mu_J' \\ \tau_J' \\ \end{array}
\right),\qquad
$$
where $J \in \{ L,\ R \}$.
We write the Standard
Model fermionic electroweak doublets as 
$$
{\bf Q_L'}_i=\left( \begin{array}{c} {\bf u_L'}_i \\ {\bf d_L'}_i \end{array}
\right),\qquad
{\bf L_L'}_i=\left( \begin{array}{c} {\bf n_L'}_i \\ {\bf e_L'}_i \end{array} \right).
$$
In order to change the na\"i{ve} model to respect $SU(2)_L$, we begin by
defining couplings to the fermionic electroweak doublet fields
\begin{equation}
\mathcal{L}_{Z^\prime f} =\left( 
\overline{\bf Q_L'}_i \lambda^{(Q)}_{ij} \gamma^\rho {\bf Q_L'}_j +
\overline{\bf L_L'}_i \lambda^{(L)}_{ij} \gamma^\rho {\bf L_L'}_j
\right)Z^\prime_{\rho},  
\label{firstSU2}
  \end{equation}
where we have used the Einstein summation convention over the family indices
$i,j \in \{1,\ 2,\ 3\}$ as well as
over the vector index $\rho$ but we have omitted gauge labels.
$\lambda^{(Q)}$ and $\lambda^{(L)}$ are Hermitian
dimensionless 3 by 3 matrices of coupling constants. Their structure will be
decided by the $Z^\prime$ ultra-violet completion (for example they will be diagonal
if it derives from an abelian group). For now we remain agnostic as to their
structure, in the spirit of simplified model building. We shall later fix them
 to give simple couplings in the mass eigenbasis. 

In order to do this, we write the terms of the Lagrangian leading to fermion
masses as 
\begin{equation}
-\mathcal{L}_{Y}=\overline{\bf Q'_L}  Y_u \phi^c {\bf u'_R} +
\overline{\bf Q'_L}  Y_d \phi  {\bf d'_R} +
\overline{\bf L'_L}  Y_e \phi  {\bf e'_R} + \frac{1}{2}
({{\bf L'_L}^T} \phi)  M^{-1}  ({\bf L'_L}\phi) + h.c. \label{yuk}
\end{equation}
where $\phi$ is the SM Higgs doublet\footnote{$\phi^c$ denotes
  $({\phi^0}^\ast,\ -{\phi^+}^\ast)^T$.} and
$Y_{u},Y_d,Y_e$ are 3 by 3 complex dimensionless Dirac mass matrices for the
up-type quarks, 
the down-type quarks and the charged leptons, respectively and
$M^{-1}$ is a 3 by 3 complex symmetric matrix of mass dimension -1. The last term
is a dimension 5 non-renormalisable operator that yields left-handed
Majorana neutrino masses. 
It may result from
integrating out heavy right-handed neutrinos, or lepton number violating
sparticles, for example. 
After electroweak symmetry breaking, the terms in Eq.~\ref{yuk} become the
fermion mass terms plus some Higgs interactions:
\begin{eqnarray}
-\mathcal{L}_{Y}&=&\overline{\bf u'_L} V_{u_L} V_{u_L}^\dagger m_u V_{u_R}
V_{u_R}^\dagger {\bf u'_R} +
\overline{\bf d'_L} V_{d_L} V_{d_L}^\dagger m_d  V_{d_R} 
V_{d_R}^\dagger {\bf d'_R} + \nonumber \\ &&
\overline{\bf e'_L} V_{e_L} V_{e_L}^\dagger Y_e  V_{e_R} 
V_{e_R}^\dagger {\bf e'_R} + 
\overline{{\bf n'_L}^c}  V_{{\nu}_L}^* V_{{\nu}_L}^T
                                            m_\nu  V_{{\nu}_L} 
V_{{\nu}_L}^\dagger {\bf n'_L} + h.c. +\ldots
\end{eqnarray}
where $V_{X_L}$ and $V_{X_R}$ are 3 by 3 unitary matrices, ${\bf n'_L}^c$ is
the charge conjugate of the left-handed neutrino field, $m_u=v Y_u$, $m_d=v
Y_d$, $m_e=v Y_e$ and $m_\nu = v^2 M^{-1}$. $v$ is the vacuum expectation
value of the neutral component of $\phi$.

Choosing
$V_{X_L}^\dagger m_X  V_{X_R}$ to be diagonal, real and positive for $X
\in \{ u,d,e\}$ and
$V_{{\nu}_L}^T m_\nu  V_{{\nu}_L}$ to be diagonal, real and positive for the
neutrinos
(all in increasing order of mass
toward the bottom right of the matrix), we can identify the {\em non}\/-primed mass
eigenstates
\begin{eqnarray}
{\bf u_R}\equiv V_{u_R}^\dagger {\bf u_R}', \qquad &
{\bf u_L}\equiv V_{u_L}^\dagger {\bf u_L}', \qquad &
{\bf d_R}\equiv V_{d_R}^\dagger {\bf d_R}', \qquad
{\bf d_L}\equiv V_{d_L}^\dagger {\bf d_L}',  \nonumber \\
{\bf e_R}\equiv V_{e_R}^\dagger {\bf e_R}', \qquad &
{\bf e_L}\equiv V_{e_L}^\dagger {\bf e_L}', \qquad &
{\bf n_L}\equiv V_{\nu_L}^\dagger {\bf n_L}'. \nonumber 
\end{eqnarray} 
We may then identify the Cabibbo-Kobayashi-Maskawa matrix $V$ and the
Pontecorvo-Maki-Nakagawa-Sakata matrix $U$:
\begin{equation}
V=V_{u_L}^\dagger V_{d_L}, \qquad U = V_{\nu_L}^\dagger V_{e_L}.
\end{equation}
Then Eq.~\ref{firstSU2} becomes 
\begin{equation}
\mathcal{L} =\left( 
\overline{\bf u_L} V \Lambda^{(Q)} V^\dagger \gamma^\rho {\bf u_L} + 
\overline{\bf d_L} \Lambda^{(Q)} \gamma^\rho {\bf d_L}
+ 
\overline{\bf n_L} U \Lambda^{(L)} U^\dagger \gamma^\rho {\bf n_L} + 
\overline{\bf e_L} \Lambda^{(L)} \gamma^\rho {\bf e_L}\right)Z^\prime_{\rho},  
\label{secSU2}
  \end{equation}
where we have defined the 3 by 3 dimensionless coupling matrices
\begin{equation}
\Lambda^{(Q)} \equiv V_{d_L}^\dagger \lambda^{(Q)} V_{d_L} , \qquad
\Lambda^{(L)} \equiv V_{e_L}^\dagger \lambda^{(L)} V_{e_L}. \label{lambdas}
\end{equation}

\subsection{The `mixed up-muon' (MUM) model}
In order to obtain the couplings in Eq.~\ref{wrongNaive}, 
 we set
\begin{equation}
\Lambda^{(Q)} = g_{bs} \left( \begin{array}{ccc} 
0 & 0 & 0 \\
0 & 0 & 1 \\
0 & 1 & 0 \\ \end{array}
\right), \qquad
\Lambda^{(L)} = g_{\mu\mu} \left( \begin{array}{ccc} 
0 & 0 & 0 \\
0 & 1 & 0 \\
0 & 0 & 0 \\ \end{array}
\right), \label{naiveCouplings}
\end{equation}
where in the ultra-violet completion we may expect $g_{bs}$ and $g_{\mu\mu}$
to be related in some way, but in our simplified model we leave them free and to be
determined by data. By the choice in Eq.~\ref{naiveCouplings}, we retain the
desired $Z^\prime$ couplings in the 
down quarks and charged leptons of 
Eq.~\ref{wrongNaive}, but these come with $SU(2)_L$-respecting mixed
couplings to up quarks and neutrinos. From now on, we refer to
Eqs.~\ref{secSU2},\ref{naiveCouplings} as the `mixed-up muon' model.
The inclusion of neutrinos into the model means that the $Z^\prime$ has a lower
BR into muons than the na\"{i}ve model: the $Z^\prime$ BR to muon pairs is
identical to that into neutrinos, to a very good approximation.

\subsection{The `mixed down-muon' (MDM) model}
Here, we simply make a different choice for $\Lambda^{(Q)}$, but the same
choice as MUM for $\Lambda^{(L)}$:
\begin{equation}
\Lambda^{(Q)} = g_{tt} V^\dagger \cdot \left( \begin{array}{ccc} 
0 & 0 & 0 \\
0 & 0 & 0 \\
0 & 0 & 1 \\ \end{array}
\right) \cdot V, \qquad
\Lambda^{(L)} = g_{\mu\mu} \left( \begin{array}{ccc} 
0 & 0 & 0 \\
0 & 1 & 0 \\
0 & 0 & 0 \\ \end{array}
\right), \label{mdmCoup}\end{equation}
This is just a rewriting of the 33$\mu\mu$ model in
Ref.~\cite{Allanach:2017bta}, but we call it the `mixed-down muon' model
as it better fits with our chosen nomenclature above.
The MDM model differs from the MUM model in that the $Z^\prime$ has
various couplings to mixed down-type quarks (and to the left-handed top). It
thus constitutes a different case for study. Matching
$\Lambda^{(Q)}$ here with Eq.~\ref{wrongNaive} identifies
\begin{equation}
g_{bs}=V_{ts}^\ast V_{tb} g_{tt}. \label{gsb}
\end{equation}
$g_{tt}>0$ ensures $g_{bs}<0$ as required by Eq.~\ref{constraint}, since 
$V_{ts}\approx -0.04$ and $V_{tb} \approx 1$.
Since $|V_{ts}^\ast V_{tb}| \approx 0.04 \ll 1$, this model helps explain why
the $Z^\prime$ model couples more weakly to $\overline{b_L} s_L + h.c.$ as
compared to $\overline{\mu_L} \mu_L$, making the $B_s - \overline{B_s}$ mixing
constraint in Eq.~\ref{bsmix} easier to satisfy at the same time as
Eq.~\ref{constraint}. 
A more complete MDM type model is provided 
by the Third
Family Hypercharge Model example case~\cite{Allanach:2018lvl}, which predicts that
 the relevant couplings have similar structure to those in
Eq.~\ref{mdmCoup} (along with other couplings to third generation fermions and
mixed neutrinos and small violations of flavour universality in the $Z$
couplings).  

Both the MDM and the MUM model generate the Feynman diagrams for hadron collider
di-muon production shown in Fig.~\ref{fig:feyn} (along with additional
production diagrams from other quarks). 

\subsection{Decays}
The $Z^\prime$ partial width for decays into 
massless fermions $f_i$ and $\overline{f_j}$ is given by 
\begin{equation}
  \Gamma_{f_i f_j}  = \frac{C}{24\pi} |g_{f_if_j}|^2 M_{Z^\prime} \, ,
  \label{eq:width}
  \end{equation}
where the constant $C=3$ for coloured fermions and $C=1$ for colour singlet
fermions. In the models we study (defined by the couplings in
Eqs.~\ref{naiveCouplings},\ref{mdmCoup} and Eq.~\ref{secSU2}) we
sum over 
decays to all of the SM fermions that the $Z^\prime$ couples to. Given the
requirements on 
the couplings to fit the flavour anomalies, the decay width is dominated by
decays to muons and neutrinos for both the MUM and the MDM models. 
In the limit that the fermion masses are small compared to $M_{Z^\prime}$
(this will be a good approximation in the domain of parameter space we consider),
Eq.~\ref{eq:width} implies that the BRs are independent of
$M_{Z^\prime}$. 

\begin{figure}
  \begin{center}
\unitlength=15cm
\begin{picture}(1,0.5)(0,0)
    \put(0,0){\includegraphics[width=0.5\textwidth]{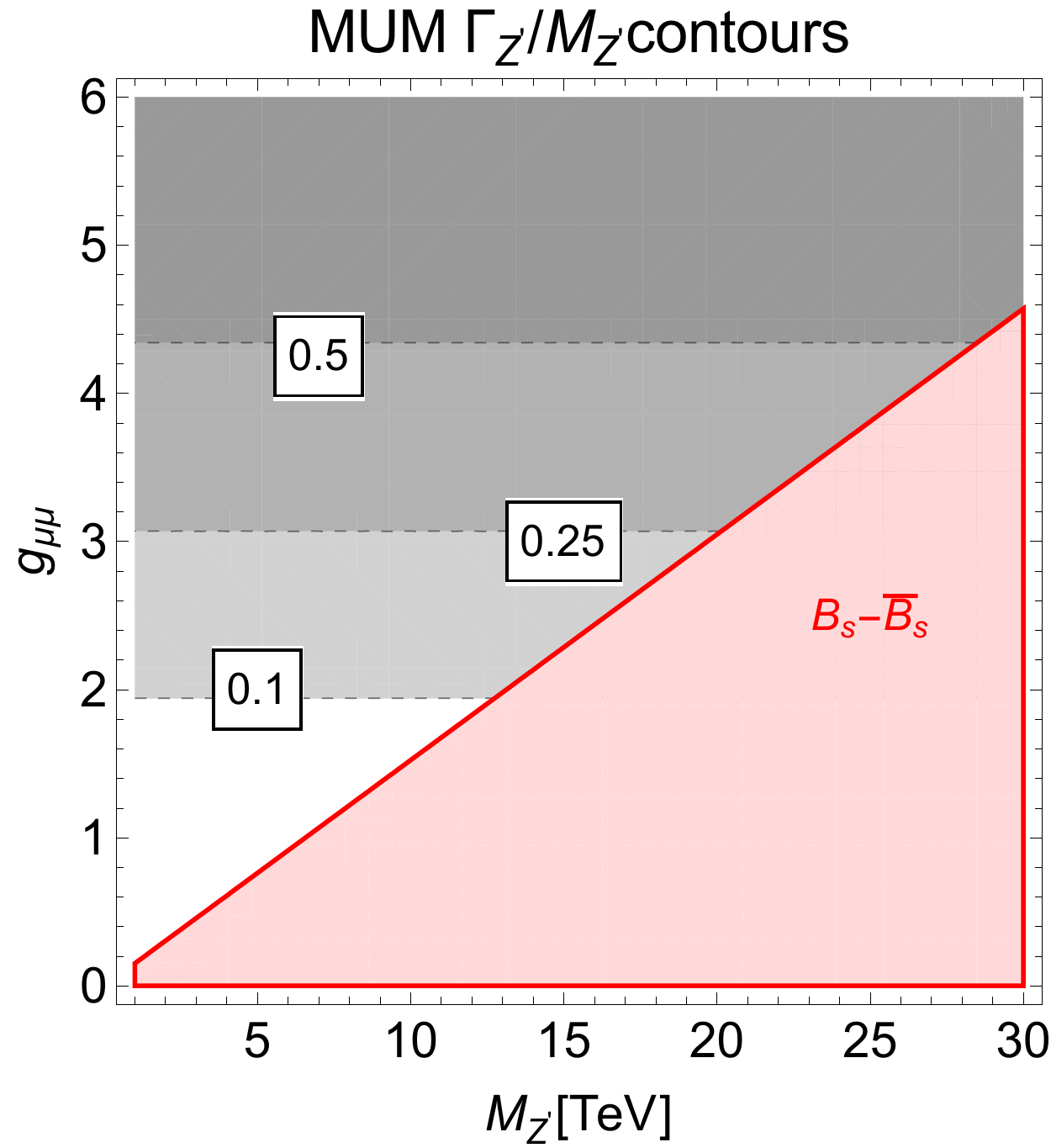}}
    \put(0.5,0){\includegraphics[width=0.5\textwidth]{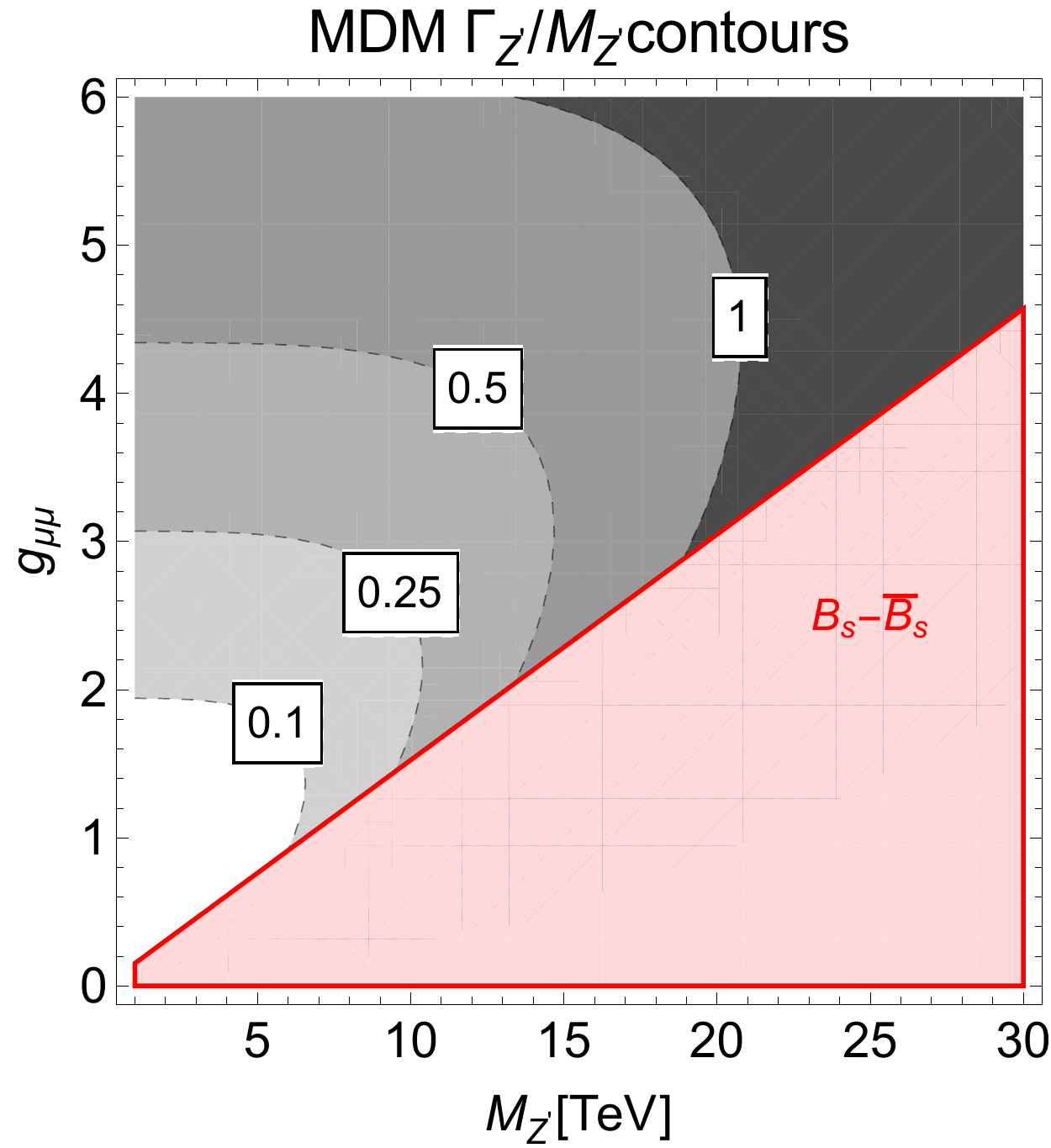}}
\put(0.1,0.085){\sf \footnotesize Allanach, Corbett, Dolan, You, 2018}
\put(0.6,0.085){\sf \footnotesize Allanach, Corbett, Dolan, You, 2018}
\end{picture}
\end{center}
\caption{\label{fig:gbs} Width of the $Z^\prime$ as a
  fraction of its mass as a function of $M_{Z^\prime}$ and $g_{\mu\mu}$ in the
  MUM (left hand panel) and MDM (right hand panel) models, derived from
  Eq.~\ref{eq:width}. 
Each point in the parameter plane fits the neutral current
  $B-$anomalies since $g_{bs}$ has been chosen to satisfy 
  Eq.~\ref{constraint} with $x=1.00$. 
The black region at the top right hand panel has width greater or equal to the
mass, meaning that the model has entered the non-perturbative r\'{e}gime, and
our results (based on perturbation theory) may be inaccurate there.} 
\end{figure}

We show in Fig~\ref{fig:gbs} the width $\Gamma$ as a function of
$M_{Z^\prime}$ and $g_{\mu\mu}$ for the
MUM (left panel) and MDM (right panel) models, with the region in
red ruled out by the constraint from $B_s$-mixing. This forces 
$|g_{bs}|$ to be small, except when $M_{Z^\prime}$ is large. We also
note that the relative width increases rapidly with $M_{Z'}$ in the MDM
model. This is because in the MDM model, as $M_{Z^\prime}$ becomes large,
$g_{tt}$ is driven to be large by Eqs.~\ref{gsb} and~\ref{constraint}. Because the
$\bar b s$ coupling is unsuppressed by a CKM mixing element in the MUM model,
we do not see the effect there.
In our simplified models, it could also be sensible to add additional
couplings of the $Z^\prime$ (these are often present in specific models). 
In that case, one could consider the width to be a free parameter with a
minimum value given by the simplified model value shown in Fig.~\ref{fig:gbs}.
A larger relative width, if it is larger than the
experimental resolution,  typically means that the
sensitivity is reduced and searches are consequently more challenging.

In the MUM model, Eq.~\ref{relstr} means that the $Z^\prime$ decays with a $50\%$
BR to muon pairs and a $50\%$ BR to neutrinos, to a
good approximation. However, in the MDM model, $|g_{tt}| = |g_{bs}| /
|V_{ts}^* V_{tb}|$ enhances the coupling to quarks: putting in the central values
$|V_{ts}|=0.04$, $|V_{tb}|=1$ for the magnitudes of CKM matrix elements yields
$y \equiv |g_{tt}| / |g_{\mu\mu}| < 0.6$. The BR into quarks in the MDM
model is then approximately 
\begin{equation}
z \equiv \sum_{i,j=1}^3 BR(Z^\prime \rightarrow q_i \bar q_j) = \frac{3 y^2}{1 + y^2},
\label{zdef}
\end{equation}
which can be as high as 52$\%$. 
The remainder of the decays are again split
equally between muon pairs and neutrino pairs.
\begin{table}\begin{center}
\begin{tabular}{|cc|cccccccc|} \hline
\multicolumn{2}{|c|}{MUM model} & \multicolumn{8}{|c|}{MDM model} \\ \hline
mode & BR & mode & BR & mode & BR& mode &BR & mode & BR\\
$\nu_i \bar \nu_k$ & 0.5 & $\nu_i \bar \nu_k$& $(1-z)/2$ & $t \bar t$& $z/2$
& $j j^\prime$ &$y^2zX/2$ & $b j$ & $y^2zY/2$\\
$\mu^+ \mu^-$ & 0.5 & $\mu^+ \mu^-$ & $(1-z)/2$ &$\bar b b$&
      $y^2z|V_{tb}|^4/2$
                             & $\bar b j$& $y^2zY/2$& & \\
\hline\end{tabular}
\caption{Summary of BRs (BRs) of $Z^\prime$ for the MUM and MDM
  models. We have categorised the three lighter quarks and anti-quarks  into a
  generic 
  `light' jet $j, j^\prime \in \{ u,d,s,\bar u, \bar
  d, \bar s\}$. $i$ and $k$ $\in \{1,2,3\}$ are
  family indices, $X\equiv\left| |V_{td}|^2 + |V_{ts}|^2 + 2 \Re(V_{ts}^\ast
                 V_{td}) \right|^2$,
$Y=|V_{tb}|^2 |V_{td} + V_{ts}|^2$,
$y=|g_{tt}|/|g_{\mu\mu}|$ and $z$ is defined in Eq.~\ref{zdef}.
\protect\label{zprimebrs} }
\end{center}
\end{table}
We summarise the BRs for both the MUM model and the MDM model in
Table~\ref{zprimebrs}. Since it is difficult or impossible to discriminate the
flavour of light jets, we have lumped them all together.
The table already suggests channels to search for the $Z^\prime$. Muon
anti-muon pairs have a sizeable BR in any event, and will be the
primary search channel, being the most closely related channel to the
explanation of the neutral current $B-$anomalies.
It is this channel that we shall focus on in the
present paper. However, in the MDM model, a sizeable BR to
boosted top anti-top pairs is also possible, and the resulting boosted top
pairs are an interesting channel for future study.

\subsection{Indirect sensitivity in high invariant mass di-muon tails}

The $Z^\prime$ may be too heavy to be directly produced on-shell at the
LHC\@. Nevertheless, it could still leave an indirect imprint in the high
invariant mass
di-muon tail at the LHC~\cite{Greljo:2017vvb,Alioli:2017nzr}. In this case we
may use an EFT approach in which we need only consider the four-fermion
operators induced by integrating out the $Z^\prime$. This can give an additional
signal contribution to the di-muon final state above the usual SM Drell-Yan
background.

We use the ATLAS data at 13 TeV with 36.1 fb$^{-1}$ with the observed and SM
number of events per bin given by Table 9 of Ref.~\cite{Aaboud:2017buh}.
Following Ref.~\cite{Greljo:2017vvb}, we parameterise the expected number of
events per bin as a function of the four-fermion operator coefficients
$C_{q\bar{q}}$ by  
\begin{equation}
\frac{N_{\text{bin}}}{N_{\text{bin}}^{\text{SM}}} = \frac{\sum_{q, \bar{q}}
  \int_{\tau_{\text{min}}}^{{\tau_{\text{max}}}} d\tau \tau L_{q\bar{q}}(\tau,
  \mu_F) |F_{q\bar{q}}(\tau s, C_{q\bar{q}})|^2 }{\sum_{q_i, \bar{q}_j}
  \int_{\tau_{\text{min}}}^{{\tau_{\text{max}}}} d\tau \tau L_{q\bar{q}}(\tau,
  \mu_F) |F_{q\bar{q}}^{\text{SM}}(\tau s)|^2}.
\end{equation}
Thus the new physics effects are only taken into account via the EFT
approximation, or the signal is not sensitive to the shape of the $Z^\prime$
propagator. This is an approximation which is only valid when $\hat s \ll
M_{Z^\prime}^2$, and we must take care to delineate the domain of its validity.
The sum is over all five parton flavours $q \in \{ u,c,d,s,b\}$ and the parton
luminosity function 
can be written as 
\begin{equation}
L_{q\bar{q}}(\tau, \mu_F) = \int_\tau^1 \frac{dx}{x} f_q(x, \mu_F) f_{\bar{q}}(\tau/x, \mu_F) \, ,
\end{equation}
where $f$ is the parton distribution function and $\mu_F$ the factorisation
scale that we set to $\hat{s}$; we also defined $\tau \equiv \hat{s} / s$ with $s$ the proton-proton
squared centre of mass energy and $\hat{s} \equiv M_{\mu\mu}^2$. The
propagator function $F$ is given by the expression 
\begin{equation}
F_{ij}(p^2, C_{ij}) = \delta^{ij}\left( \frac{e^2 Q_q Q_l}{p^2} + \frac{g_Z^q g_Z^l}{p^2 - M_Z^2 + i M_Z \Gamma_Z} \right) + \frac{C_{ij}}{v^2}.
\end{equation}
$Q_{f}$ is the electric charge of the fermion, and $g_Z^f \equiv
\frac{2M_Z}{v}\left( T_f^3 - Q_f \sin^2{\theta_W} \right)$ where $f$ label the
species of fermion,
$\theta_w$ is the SM Weinberg angle and $T_f^3$ is the diagonal generator of
$SU(2)$. 
\begin{figure}
\centering
 \includegraphics[width=0.55 \textwidth]{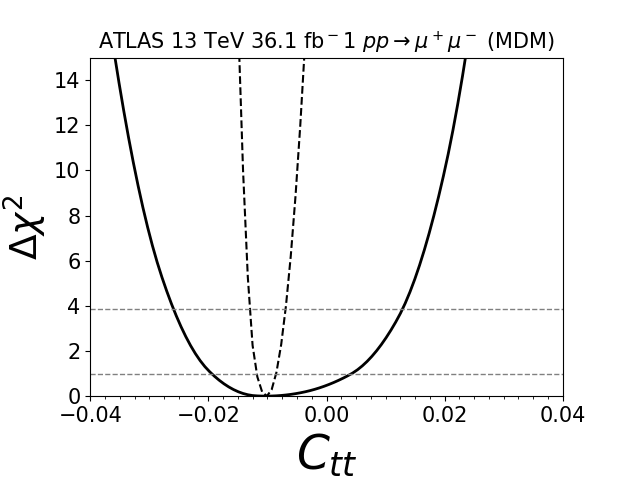}
\caption{$\Delta \chi^2$ for the four-fermion coefficient $C_{tt} \simeq C_{bb} = - g_{bb}g_{\mu\mu}v^2/M_{Z^\prime}^2$ from a fit to the ATLAS 13 TeV 36.1 fb$^{-1}$ di-muon distribution in the MDM model. The high-luminosity LHC projections are shown in dashed lines.  }
\label{fig:MDMcbbchisquared}
\end{figure}

There was no region of 95$\%$ CL sensitivity of this
high di-muon invariant mass tail analysis to the MUM model, whose signal
cross-section is too small at $\sqrt{s}=13$ TeV,
so we here focus on the sensitivity of the MDM model.
Treating each bin as independently Poisson-distributed, we perform a $\chi^2$
fit for the 
MDM model in which the coupling to $b\bar{b}$ is allowed to vary
freely with the muon coupling fixed to $g_{\mu\mu} = 1.5$. The couplings to
the other flavours are dependent on $g_{bb} \equiv g_{tt}/|V_{tb}|^2 \approx g_{tt}$
through CKM rotations, \`{a} la Eq.~\ref{mdmCoup}. The 
resulting $\Delta \chi^2$ for the four-fermion operator coefficient $C_{tt} \simeq C_{bb} =
-g_{bb}g_{\mu\mu}v^2/M_{Z^\prime}^2$ is shown in 
Fig.~\ref{fig:MDMcbbchisquared}\footnote{Switching on one operator
  coefficient at a time, we find limits in good agreement with those
  in Ref.~\cite{Greljo:2017vvb}.}. 

\begin{figure}
\begin{center}
\unitlength=15cm
\begin{picture}(0.5,0.5)(0,0)
\put(0,0){\includegraphics[width=0.45 \textwidth]{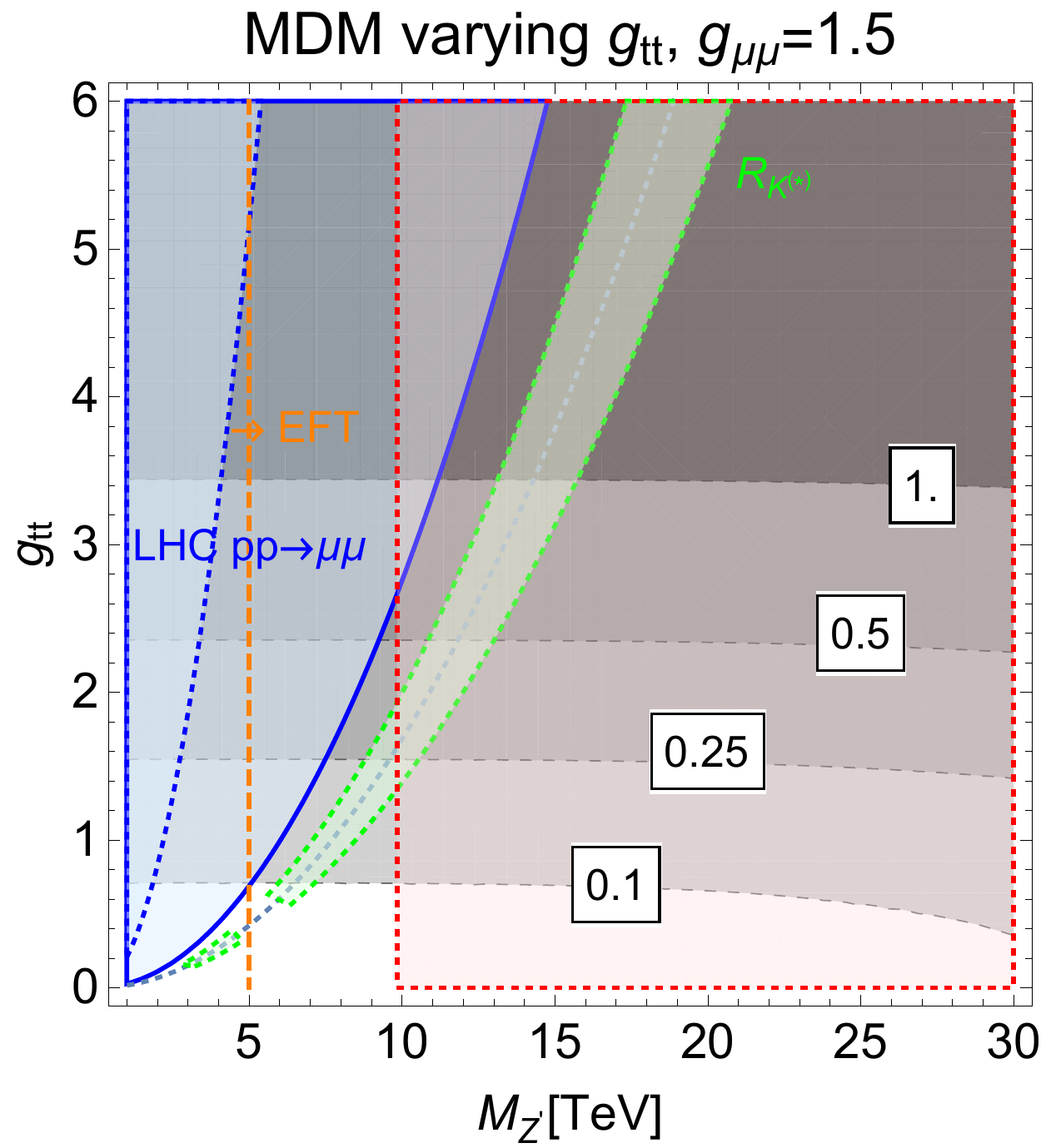}}
\put(0.42,0.12){\rotatebox{90}{\sf \footnotesize Allanach, Corbett, Dolan, You, 2018}}
\end{picture}
\caption{95\% CL exclusion region of the ATLAS 13 TeV $p p \to \mu \mu$ di-muon
  distribution (blue region) for 36.1 fb$^{-1}$ (3 ab$^{-1}$) denoted by dotted (solid) blue lines, as a function of $M_{Z^\prime}$ vs $g_{tt}\approx
  g_{bb}$ for
  a deformed MDM model with $g_{\mu\mu}=1.5$. $g_{bs}$ is fixed to its anomaly
  compatible value. 
The region of EFT validity for this analysis
  lies to the right of the vertical orange dotted line. Contours of
  $\Gamma_{Z^\prime}/M_{Z^\prime}$ are represented by dashed grey lines. The
  anomaly-compatible region within 1$\sigma$ is shown in green, while the
  $B_s-\overline{B_S}$ mixing constraint is in red.}
\label{fig:MDMdimuontail}
\end{center}
\end{figure}

In Fig.~\ref{fig:MDMdimuontail}, the 95\% CL exclusion region of the di-muon mass
distribution to a deformed MDM model is plotted as the blue region on the parameter
space of $M_{Z^\prime}$ vs $g_{bb}$ bounded by the dotted blue curve. 
The deformation we consider is the removal of the connection between $g_{bs}$
and $g_{tt}$ in Eq.~\ref{gsb}: they are now considered to be independent.
The 
dashed (solid) blue curve shows expected sensitivity for the (HL-)LHC\@.
The vertical dashed orange line
represents (approximately) where the $Z^\prime$ mass is 
beyond the direct reach of the LHC such that
the EFT approach is valid. The horizontal dashed grey lines labelled by white
boxes represent the width as a fraction of the $Z^\prime$ mass. 
Thus, we deduce that the blue region to the right hand side of the EFT line is
ruled out by ATLAS\@. 
Should a sizeable deviation appear in the di-muon tail at the LHC (but should no
resonance appear), this will be somewhere to the right-hand side of the
current (dashed blue) exclusion region on the plot. 
In the MDM
model, this along with EFT validity necessarily points towards a wide
$Z^\prime$ ($\Gamma/M_{Z^\prime} 
\geq 0.1$) to be searched for at
future higher-energy colliders. 
Note that this conclusion is more general than
the specific case of an anomaly-compatible $Z^\prime$, whose parameter space
would then have to lie within the green band. The discovery of
indirect effects that may still show up at the high-luminosity LHC therefore
would provide additional motivation for studying future sensitivities to large
width resonances.

\section{Direct $Z^\prime$ Sensitivity of Hadron Colliders}
\label{sec:sens}

We shall here focus on the $\mu^+ \mu^-$ channel for identifying fat
flavourful $Z^\prime$ 
production. The $\mu^+ \mu^-$ channel has the benefit of being directly
involved with the inferred new physics contribution in $R_{K^{(\ast)}}$, and we
know that its coupling $g_{\mu\mu}$ to the $Z^\prime$ must be much larger than
$|g_{bs}|$  because of Eq.~\ref{relstr}. In particular models, it could be that
other channels may be even more sensitive (for example to boosted top pairs),
but we still restrict ourselves to the $\mu^+\mu^-$ channel because of its
omnipresence in models which explain the $R_{K^{(\ast)}}$ measurements.

\subsection{Methodology}
\label{sec:meth}
Aside from resonance searches in the narrow width approximation~\cite{Allanach:2017bta},
some
previous work on the collider prospects for $Z^\prime$s which explain the flavour
anomalies has focused on precision measurements of the high invariant mass
tails of 
di-lepton distributions at the LHC~\cite{Greljo:2017vvb,Alioli:2017nzr}. 
Other studies have focused on production of the resonance in association
with a $b$-jet, exploiting the flavour structure of the $Z^\prime$
couplings~\cite{Afik:2018nlr,Kohda:2018xbc}. Our
strategy is instead a direct search for the fat $Z^\prime$ resonance in the
di-lepton 
invariant mass distribution, taking width effects correctly into
account. 
We use the 5-flavour {\tt
  NNPDF2.3LO}~\cite{Ball:2012cx} $(\alpha_s(M_Z)=0.119)$ parton distribution
functions 
via {\tt LHAPDF6}~\cite{Buckley:2014ana} in 
order to re-sum the logarithms associated with the initial state
$b$-quark~\cite{Lim:2016wjo}. The 
hadronisation scale is fixed to be $M_{Z^\prime}$.

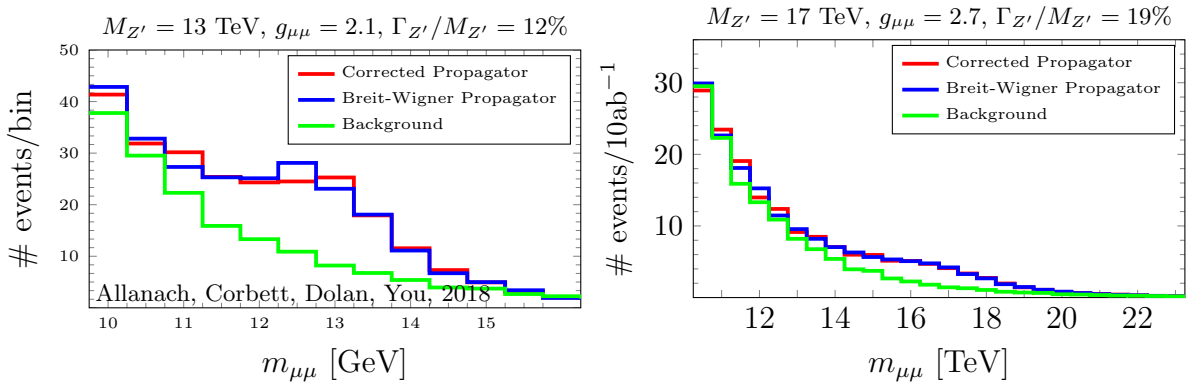
\begin{figure}
\begin{tabular}{c}
\begin{tikzpicture} \begin{axis}
[
    title={$M_{Z^\prime}=13$ TeV, $g_{\mu\mu}=2.1$, $\Gamma_{Z^\prime}/M_{Z^\prime}=12\%$},
    tiny,
    minor y tick num = 5,
    minor x tick num = 3,
    ymin = 0,ymax = 50,
    xmin = 9.750,xmax = 16.250,
    xlabel={\large{$m_{\mu\mu}$ [GeV]}},
   xtick = {10,11,12,13,14,15},
   ylabel={\large{$\#$ events/bin}},
    ytick = {0,10,20,30,40,50},
    yticklabels = {$$,$10$,$20$,$30$,$40$,$50$},
     legend style={
        cells={anchor=west},
    },
    legend entries={Corrected Propagator,Breit-Wigner Propagator,Background},
    width=1.61*5cm,
    height=5cm
        ]

    \addplot [
        const plot, draw=red, line width=0.05cm
    ] coordinates {
        (9.75,41.35459)
        (10.25,31.86249)
        (10.75,30.164369999999995)
        (11.25,25.39171)
        (11.75,24.3209)
        (12.25,24.50577)
        (12.75,25.27766)
        (13.25,17.925)
        (13.75,11.549769999999999)
        (14.25,7.319485)
        (14.75,4.9291339999999995)
        (15.25,3.0631099999999996)
        (15.75,1.993434)
        (16.25,1.993434)
    };
    
    \addplot [    
            const plot, draw=blue, line width=0.05cm
    ] coordinates {
        (9.75,42.83712999999999)
        (10.25,32.80938)
        (10.75,27.318169999999995)
        (11.25,25.306079999999998)
        (11.75,25.124879999999997)
        (12.25,28.10171)
        (12.75,23.089789999999997)
        (13.25,18.07576)
        (13.75,11.106480000000001)
        (14.25,6.706305999999999)
        (14.75,4.98224)
        (15.25,3.4072249999999995)
        (15.75,1.917565)
        (16.25,1.917565)
    };
    
                \addplot [
        const plot, draw=green, line width=0.05cm
    ] coordinates {

        (9.75,37.77578999999999)
        (10.25,29.51798)
        (10.75,22.298309999999997)
        (11.25,15.885979999999998)
        (11.75,13.31181)
        (12.25,10.889009999999999)
        (12.75,8.201153)
        (13.25,6.754377999999999)
        (13.75,5.396253)
        (14.25,3.943058)
        (14.75,3.729748)
        (15.25,2.669094)
        (15.75,2.248211)
        (16.25,1.798276)
    };
\node at (270,25) {\footnotesize Allanach, Corbett, Dolan, You, 2018};    
    \end{axis}

\end{tikzpicture}
\begin{tikzpicture} \begin{axis}
[   title={$M_{Z^\prime}=17$ TeV, $g_{\mu\mu}=2.7$, $\Gamma_{Z^\prime}/M_{Z^\prime}=19\%$},
    tiny,
    minor y tick num = 4,
    ymin = 0,ymax = 36,
    xmin = 10.250,xmax = 23.250,
    xlabel={\large{$m_{\mu\mu}$ [TeV]}},
   xtick = {10.0,11.,12.,13.,14.,15.,16.,17.,18.,19.,20.,21.,22.,23.},
   xticklabels = {\large{$10$},,\large{$12$},,\large{$14$},,\large{$16$},,\large{$18$},,\large{$20$},,\large{$22$},},
   ylabel={\normalsize{$\#$ events/$10\text{ab}^{-1}$}},
    ytick = {0,10,20,30},
    yticklabels = {$$,\normalsize{$10$},\normalsize{$20$},\normalsize{$30$},$40$,$50$},
     legend style={
        cells={anchor=west},
    },
    legend entries={Corrected Propagator,Breit-Wigner Propagator,Background},
    width=1.61*5cm,
    height=5cm    ]

    \addplot [
        const plot, draw=red, line width=0.05cm
    ] coordinates {
        (10.25,28.91506)
        (10.75,23.45467)
        (11.25,19.06671)
        (11.75,13.97599)
        (12.25,12.374049999999999)
        (12.75,9.15164)
        (13.25,8.482558)
        (13.75,7.058362999999999)
        (14.25,5.983429)
        (14.75,5.953022)
        (15.25,5.139329)
        (15.75,5.108174)
        (16.25,4.700746)
        (16.75,4.104885)
        (17.25,3.3631189999999997)
        (17.75,2.74386)
        (18.25,1.962157)
        (18.75,1.4261659999999998)
        (19.25,1.08452)
        (19.75,0.7498064000000001)
        (20.25,0.6323689)
        (20.75,0.4979283)
        (21.25,0.413435)
        (21.75,0.30628259999999996)
        (22.25,0.2421589)
        (22.75,0.18307649999999998)
        (23.25,0.11546609999999999)
        (23.75,0.11546609999999999)
    };
    
                \addplot [
        const plot, draw=blue, line width=0.05cm
    ] coordinates {
        (10.25,29.899559999999997)
        (10.75,22.633419999999997)
        (11.25,18.09367)
        (11.75,15.2486)
        (12.25,11.46177)
        (12.75,9.559085)
        (13.25,8.203149)
        (13.75,7.045581)
        (14.25,6.299899999999999)
        (14.75,5.66008)
        (15.25,5.323208)
        (15.75,5.082722)
        (16.25,4.778319)
        (16.75,4.215286)
        (17.25,3.2787129999999998)
        (17.75,2.679462)
        (18.25,1.866397)
        (18.75,1.473286)
        (19.25,1.088528)
        (19.75,0.8252749)
        (20.25,0.6279159000000001)
        (20.75,0.47733410000000004)
        (21.25,0.3834607)
        (21.75,0.3204955)
        (22.25,0.2390637)
        (22.75,0.17958010000000002)
        (23.25,0.1182327)
        (23.75,0.1182327)
    };
    
        \addplot [ 
        const plot, draw=green, line width=0.05cm
    ] coordinates {
        (10.25,29.51798)
        (10.75,22.298309999999997)
        (11.25,15.885979999999998)
        (11.75,13.31181)
        (12.25,10.889009999999999)
        (12.75,8.201153)
        (13.25,6.754377999999999)
        (13.75,5.396253)
        (14.25,3.943058)
        (14.75,3.729748)
        (15.25,2.669094)
        (15.75,2.248211)
        (16.25,1.798276)
        (16.75,1.428562)
        (17.25,1.302839)
        (17.75,1.061688)
        (18.25,0.8120575000000001)
        (18.75,0.7032793)
        (19.25,0.6477686)
        (19.75,0.45663190000000003)
        (20.25,0.41970660000000004)
        (20.75,0.35323499999999997)
        (21.25,0.2875027)
        (21.75,0.2592854)
        (22.25,0.2159299)
        (22.75,0.16688530000000001)
        (23.25,0.14974890000000002)
        (23.75,0.12042829999999999)
    };

\end{axis}
\end{tikzpicture}
\end{tabular}
\caption{\label{fig:di-muondistro}Expected di-muon invariant mass
  distributions at the 
  FCC for (left) $M_{Z^\prime}$=13~TeV, $g_{\mu\mu}=2.1$ and (right)
  $M_{Z^\prime}$=17~TeV, $g_{\mu\mu}=2.7$,  corresponding to widths of 12\%
  and 19\% respectively. The expected number of events per bin on the ordinate
  is for 10~ab$^{-1}$ of integrated luminosity. This figure shows the
  difference between using the {\tt MadGraph5} default propagator and the new
  propagator 
  $\sim1/(p^2-M^2-i p^2 \Gamma/M)$. The significance for $M_{z^\prime}=13$ TeV
  is 8.5 (9.7 for 
  the default propagator) summing from bin 4 (5). The significance for $M_{Z^\prime}=17$ TeV
  is 4.6 (5.6) summing from bin 9 (10). All histograms and significance calculations are
  post-detector simulation (i.e.\ {\tt DELPHES 3}). \label{fig:prop}}  
\end{figure}

The standard propagator used in resonance production has the Breit-Wigner form
\begin{equation}
\mathcal{D}_{\mu\nu}(p^2) = \frac{-i
  \eta_{\mu\nu}}{p^2-M^2_{Z^\prime}+i\Gamma_{Z^\prime}M_{Z^\prime}},
\end{equation}
where $\eta_{\mu \nu}$ is the Minkowski metric.
This results from re-summing a class of corrections to the tree-level propagator. These are related to the decay width by the optical theorem. Usually, these corrections are evaluated at a fixed scale $\hat{s}=M^2_{Z^\prime}$, in which case the propagator has the form above. However, for wide resonances
the partonic centre-of-mass energy can be sufficiently far from the pole in
the propagator that this approximation is no longer valid. A clear exposition
of this can be found in the literature on the line-shape of the
$Z^0$-boson~\cite{Bardin:1989qr}. In this case we must include the momentum
dependence, and not just evaluate the imaginary terms at the fixed scale
$\hat{s}=M^2_{Z^\prime}$. In practical terms, this amounts to replacing
$\Gamma_{Z^\prime}M_{Z^\prime}$ with
$\frac{\hat{s}}{M_{Z^\prime}^2}\Gamma_{Z^\prime}M_{Z^\prime}$, so that the
corrected propagator has the form 
\begin{equation}
\mathcal{D}_{\mu\nu}(p^2) =
\frac{-i\eta_{\mu\nu}}{p^2-M^2_{Z^\prime}+i\frac{p^2}{M_{Z^\prime}^2}\Gamma_{Z^\prime}M_{Z^\prime}}.
  \end{equation}
In practice we do this by changing the $Z^\prime$ propagator in the {\tt UFO} files~\cite{Christensen:2008py}
we generate\footnote{The {\tt UFO} files are included in the ancillary information
  submitted with the {\tt arXiv} version of this paper.} from
\texttt{FeynRules}~\cite{Degrande:2011ua,Alloul:2013bka}. We generate events using \texttt{MadGraph5}~\cite{Alwall:2014hca}. We show the effects of this change from the 
Breit-Wigner form in Fig.~\ref{fig:prop}, which shows the di-muon invariant
mass distribution for $m_{Z^\prime}=13,17$~TeV for the Breit-Wigner
and corrected
propagators. We observe a smearing due to the large width effect,
which reduces sensitivity somewhat.

We define signal sensitivity as follows: first, we define a window of di-muon
invariant mass in which to
generate events, depending upon the collider: 
\begin{eqnarray}
m_{\mu\mu}^{\text{HL-LHC}} &\in& [ \text{max}\{M_{Z^\prime}-\Gamma-500\text{~GeV},\ 100 \text{~GeV}\} ,\
\text{min}\{M_{Z^\prime} + \Gamma,\ 5.9 \text{~TeV}\} ]\nonumber\\ 
m_{\mu\mu}^{\text{HE-LHC}} &\in& [ \text{max}\{M_{Z^\prime}-\Gamma-2\text{~TeV},\
250\text{~GeV}\},\ 
\text{min}\{M_{Z^\prime}
  + \Gamma,\ 11.25\text{~TeV}\} ]\nonumber\\
m_{\mu\mu}^{\text{FCC}} &\in& [ \text{max}\{M_{Z^\prime}-\Gamma-2\text{~TeV},\
250\text{~GeV}\},\ 
\text{min}\{M_{Z^\prime}
  + \Gamma,\ 25.25\text{~TeV}\} ]. \nonumber\end{eqnarray}
We define $S_i \equiv (\sigma^{Z^\prime+SM}_i - \sigma^{SM}_i) {\mathcal L}$, where
${\mathcal L}$ 
is assumed integrated luminosity and $i \in \{1, 
\ldots , N \}$, as the 
expected number of signal events in a single bin of  width $W$ in $m_{\mu\mu}$
estimated in our simulations\footnote{Each bin
is centred on $m_{\mu\mu}=W (2 n+ 1)/2$ GeV, where  $n \in \{0, 1, 2, \ldots
\}$.}. $W=500$ GeV is taken for all simulations apart
from the HL-LHC ones, where $W=100$ GeV is taken.
$\sigma^{Z^\prime+SM}_i$ is the $pp \rightarrow \mu^+ \mu^-$ cross-section 
including the $Z^\prime$ lying in the $m_{\mu\mu}$ bin $i$ and
$\sigma^{SM}_i$ is the SM $\mu^+ \mu^-$ cross-section in the same
bin\footnote{Although interference effects between signal and background are
  automatically taken into account by {\tt MadGraph5}, in both the MUM model
  and the   MDM model, they are CKM suppressed or parton density
  function-suppressed compared to the pure signal 
  contribution.}. Each of 
these cross-sections is to be understood as being for $pp
\rightarrow \mu^+ \mu^-$ after acceptance, efficiency and detector effects. 
The total significance, measured in terms of `number of $\sigma$', is defined to be
\begin{equation}
S = \text{max}_i D_i, \qquad \text{where} \qquad D_i \equiv \frac{\sum_{j=i}^N S_j}{\sqrt{\sum_{k=i}^N
    B_k} },
\label{signif}
\end{equation}
and the number of background events in bin $k$
$B_k\equiv \sigma^{SM}_k {\mathcal L}$ is likely to be estimated  in practice by
experiments measuring control regions.  
However, there will be systematic errors and correlations involved with the
extraction of the $B_k$ which we do not take into account here. 
Our definition of signal significance is rather crude, and in the end if many
signal events are collected, shapes and other features of signal events are
likely to be used which will increase the significance. We also ignore the
effects of theoretical uncertainties (parton density function errors, higher order contributions
etc), which will tend to decrease the sensitivity.
Despite these short-comings, at this stage our crude definition of
significance 
will suffice for a reasonable but approximate estimate.

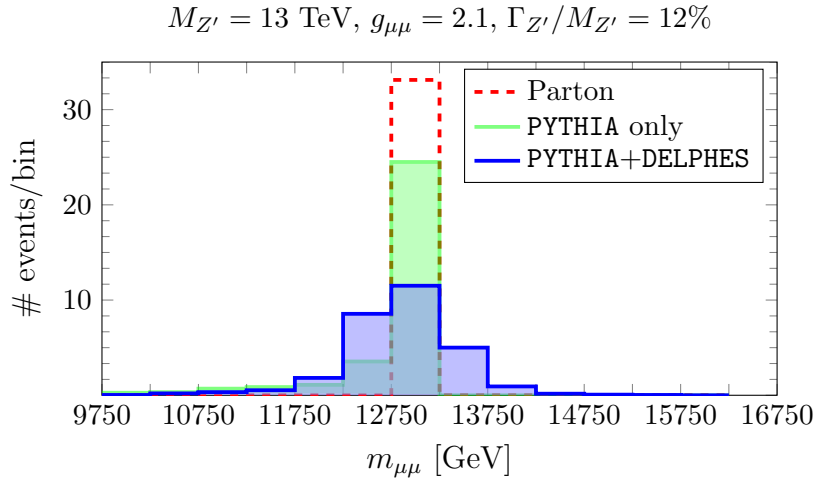
\begin{figure}
  \centering
  \begin{tabular}{c}
\begin{tikzpicture} \begin{axis}
[
    title={$M_{Z^\prime}=13$ TeV, $g_{\mu\mu}=2.1$, $\Gamma_{Z^\prime}/M_{Z^\prime}=12\%$},
    minor y tick num = 5,
    ymin = 0,ymax = 35,
    xmin = 9.750,xmax = 16.750,
    xlabel={\large{$m_{\mu\mu}$ [GeV]}},
   xtick = {9.750,10.250,10.750,11.250,11.750,12.250,12.750,13.250,13.750,14.250,14.750,15.250,15.750,16.250,16.750},
   xticklabels = {9750,,10750,,11750,,12750,,13750,,14750,,15750,,16750,,17750,},
   ylabel={\large{$\#$ events/bin}},
    ytick = {0,10,20,30,40,50},
    yticklabels = {$$,$10$,$20$,$30$,$40$,$50$},
     legend style={
        cells={anchor=west},
    },
    legend entries={Parton,{\tt PYTHIA} only,{\tt PYTHIA}+{\tt DELPHES},Background (parton)},
    width=1.61*6.5cm,
    height=6cm
        ]

        \addplot [
        const plot, draw=red, dashed, line width=0.05cm
    ] coordinates {
        (9.75,0.)
        (10.25,0.)
        (10.75,0.)
        (11.25,0.)
        (11.75,0.)
        (12.25,0.)
        (12.75,33.1212)
        (13.25,0.)
        (13.75,0.)
        (14.25,0.)
        (14.75,0.)
        (15.25,0.)
        (15.75,0.)
        (16.25,0.)
    };
    \addplot [ fill=green!50, opacity=0.5,
                const plot, draw=green, line width=0.05cm
    ] coordinates {
        (9.75,0.265234)
        (10.25,0.331543)
        (10.75,0.69624)
        (11.25,0.862012)
        (11.75,1.09409)
        (12.25,3.54751)
        (12.75,24.501)
        (13.25,0.)
        (13.75,0.)
        (14.25,0.)
        (14.75,0.)
        (15.25,0.)
        (15.75,0.)
        (16.25,0.)
    };
    
    \addplot [    fill=blue!50,fill opacity=0.5,
            const plot, draw=blue, line width=0.05cm
    ] coordinates {
        (9.75,0.0331543)
        (10.25,0.198926)
        (10.75,0.331543)
        (11.25,0.530469)
        (11.75,1.82349)
        (12.25,8.55381)
        (12.75,11.5045)
        (13.25,5.0063)
        (13.75,0.928321)
        (14.25,0.165772)
        (14.75,0.0663086)
        (15.25,0.0331543)
        (15.75,0.)
        (16.25,0.)
    };
\end{axis}
\end{tikzpicture}
\end{tabular}
\caption{The bleeding of a single parton level bin centred on $m_{\mu\mu}=13$
  TeV for the $Z^\prime$ signal at the FCC\@. At parton-level, we expect 33.1
  events (in 10 ab$^{-1}$), 
  after parton showering effects are simulated by {\tt PYTHIA} this reduces to
  31.3 and after simulating detector effects with {\tt DELPHES 3},
   29.2. \label{smear}}  
\end{figure}

In order to further examine the simulated effects of parton showering and the
detector, we simulate signal-only FCC collisions for $M_{Z^\prime}=13$ TeV. We
initially filter out all events 
other than those where the parton-level simulation gives di-muons in a single
bin centred around $m_{\mu\mu}=13$ TeV, as shown in Fig.~\ref{smear}: thus we
throw away the large width effects in this one plot for illustrative purposes
(but we include them elsewhere in this paper).
These central-bin events are then passed through {\tt
  PYTHIA8.2}~\cite{Sjostrand:2014zea,Cacciari:2011ma} in order to 
simulate initial 
state radiation and parton showering effects. We see that the initial
parton-level simulation gets smeared to lower invariant masses. However, at
such high $m_{\mu\mu}$, the muon resolution becomes significantly worse. This is because
 muon momentum is measured by the amount of  bending in the magnetic field of
the detector, but very high momentum muons will have small bending
compared to lower energy ones. Thus, simulating such detector effects is essential in order to account for
this. Here, we use the 
{\tt DELPHES 3}~\cite{deFavereau:2013fsa} fast detector simulator. We see that the 
detector smears $m_{\mu\mu}$ both to higher and lower values. Thus: the
overall effect of initial state radiation and detector effects is smearing to higher and to lower values, with a small bias
toward smaller invariant masses. Notably, this detector-level smearing was not
accounted for in Ref.~\cite{Allanach:2017bta}, which gave a cruder estimate in the
approximation that detector effects are the same in the LHC and FCC
environments. Such an approximation becomes worse for larger $M_{Z^\prime}$.
For this 13 TeV bin, some 10$\%$ of signal events are lost due to acceptance and
efficiencies.

\subsection{Results}

\begin{figure}
  \begin{center}
\unitlength=15cm
\begin{picture}(0.45,0.45)(0,0)
\put(0,0){\includegraphics[width=0.45\textwidth]{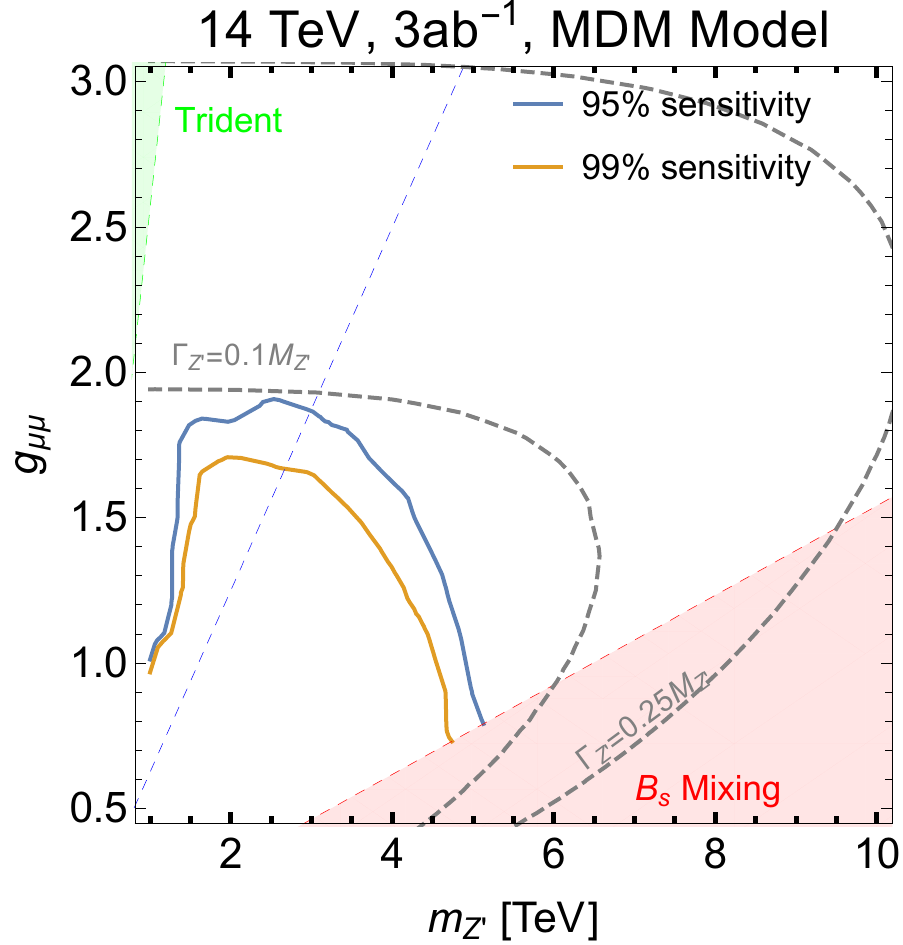}}
\put(0.424,0.072){\rotatebox{90}{\sf \footnotesize Allanach, Corbett, Dolan, You, 2018}}
  \end{picture}
\end{center}
\caption{\label{fig:sensHLLHC} Predicted sensitivity for the MDM model with 
  the 14~TeV LHC with 3/ab integrated luminosity in the
  $M_{Z^\prime}$--$g_{\mu\mu}$ plane. 
Each point in the parameter plane fits the neutral current
  $B-$anomalies since $g_{bs}$ has been chosen to satisfy 
  Eq.~\ref{constraint} with $x=1.00$. 
  The solid lines show the regions of
  95$\%$ CL and 99$\%$ CL
  sensitivity being {\em below}\/ each contour. We show
  only the MDM model, since the HL-LHC does not have sensitivity to
   the allowed MUM parameter space. The region ruled out by the $B_s$ mixing
  constraint~\cite{Arnan:2016cpy} is shown in red while the region derived
  from Ref.~\cite{DiLuzio:2017fdq} would 
  be below the blue-dashed line. The grey dashed lines show the
  values of the relative decay width $\Gamma_{Z^\prime}/M_{Z^\prime}$.
}
\end{figure}

Achieving adequate Monte-Carlo statistics in the tails of wide resonances can
be challenging. We generate events in fixed-width bins of the di-muon
invariant mass, so that we have good resolution in the tails. We find that
generating with bin widths $W=100, 500$ GeV as described in \S~\ref{sec:meth}
is sufficient to achieve smoothly 
falling distributions across the relevant range of parameter space.

The dominant background process is Drell-Yan (DY) production of di-muon pairs
via $\gamma^{*}$ and $Z$. While there are also contributions from di-boson
production, top quarks, and vector-boson plus jets, at large invariant masses
these are completely dominated by the Drell-Yan component, which makes up over
90\% of the background events at the
LHC~\cite{Aaboud:2017buh,Sirunyan:2018exx}. Accordingly, we consider DY as the
only background in our simulations.

The ATLAS di-lepton search~\cite{Aaboud:2017buh} sets limits on generic $Z^\prime$s with relative decay widths of up to 32\%, by using a mass window of twice the resonance width. However, the corresponding CMS search~\cite{Sirunyan:2018exx} only considers narrow resonances, whose widths are up to 10\% of their mass. The CMS di-jet search~\cite{Sirunyan:2018xlo} provides limits on resonances up to 30\% width, while the ATLAS di-jet searches~\cite{Aaboud:2017yvp} stay within the narrow regime.

\begin{figure}
  \begin{center}
\unitlength=15cm
\begin{picture}(1,0.5)(0,0)
\put(0,0){\includegraphics[width=0.45\textwidth]{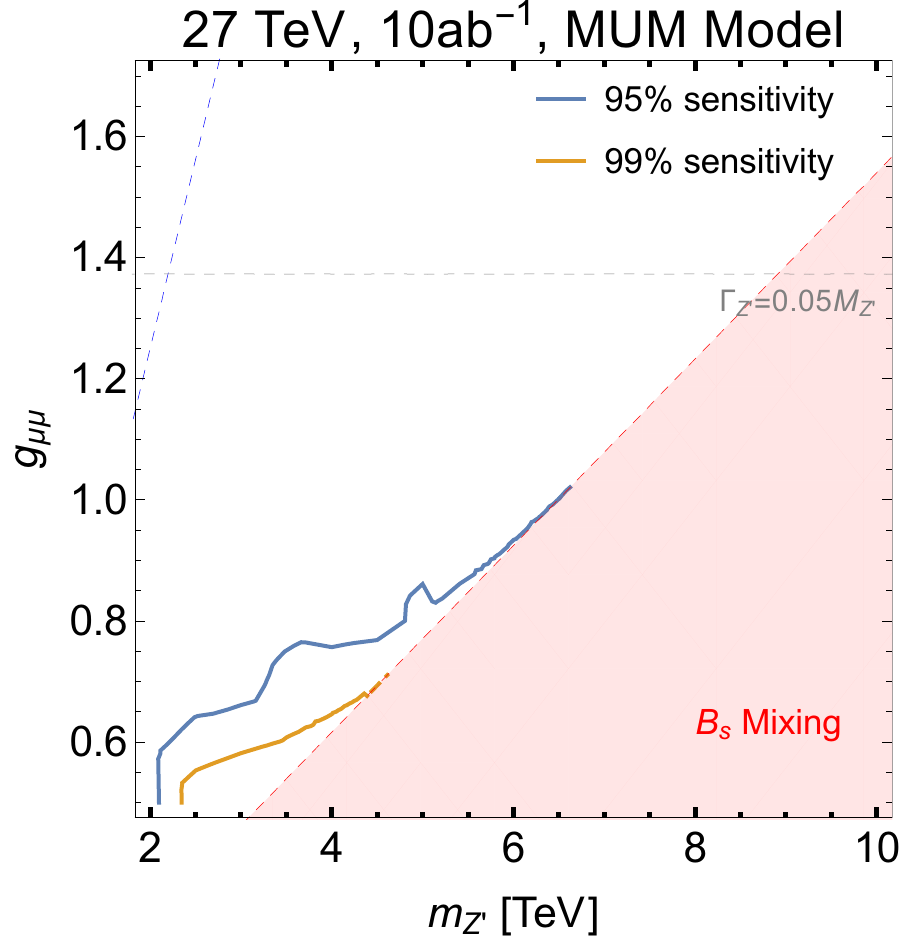}}
\put(0.5,0){\includegraphics[width=0.45\textwidth]{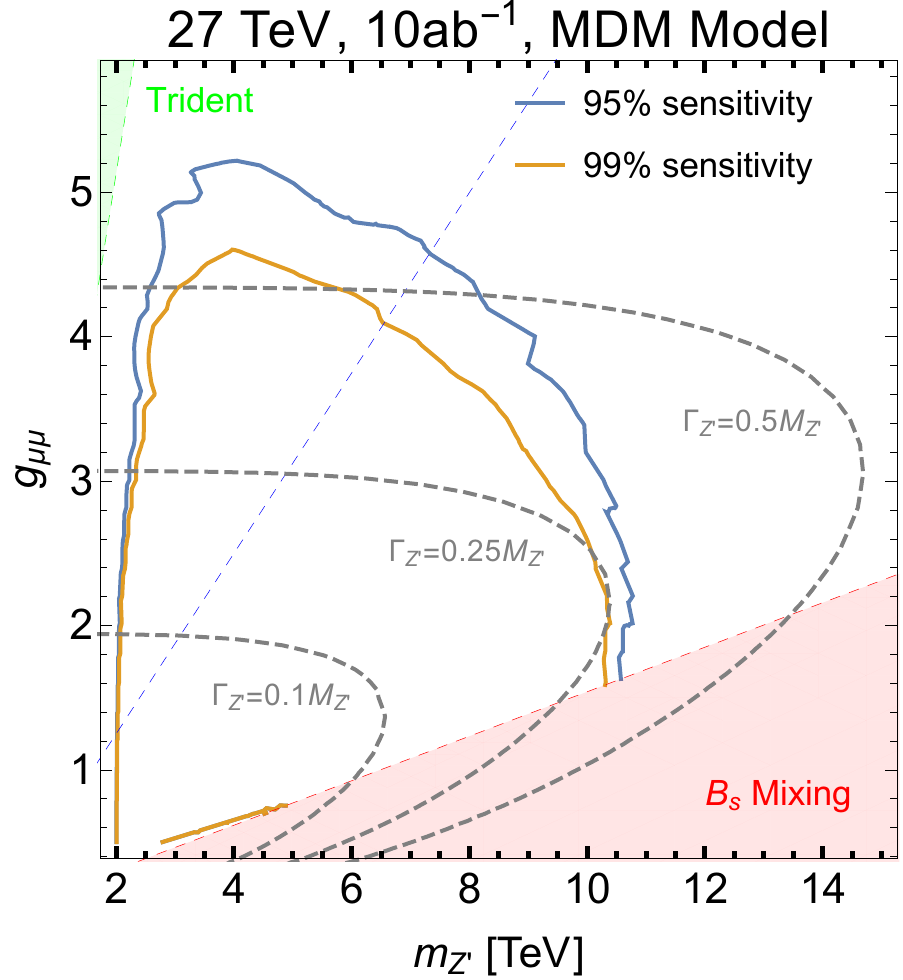}}
\put(0.12,0.067){{\sf \footnotesize Allanach, Corbett, Dolan,
    You, 2018}}
\put(0.62,0.067){{\sf \footnotesize Allanach, Corbett, Dolan, You, 2018}}
\end{picture}
\end{center}
\caption{\label{fig:sensHELHC} Predicted sensitivity for
  the 27~TeV HE-LHC with 15/ab integrated luminosity in the
  $M_{Z^\prime}$--$g_{\mu\mu}$ plane. 
  Each point in the plane fits the neutral current
  $B-$anomalies, since Eq.~\ref{constraint} with $x=1.00$ has been enforced
  meaning that $g_{bs}   \propto M_{Z^\prime}^2/g_{\mu\mu}$.
  The solid lines show the regions of
  95$\%$ CL and 99$\%$ CL sensitivity, which are {\em below}\/ each contour. The
  left-hand plot shows the MUM model, the right-hand plot the MDM model. The
  region ruled out by the $B_s$ mixing constraint~\cite{Arnan:2016cpy} is
  shown in red while the region derived from Ref.~\cite{DiLuzio:2017fdq} would
  be below the blue-dashed line. The grey dashed lines show the
  values of the relative decay width $\Gamma_{Z^\prime}/M_{Z^\prime}$.} 
\end{figure}

We find that the 14~TeV HL-LHC with $3\text{ab}^{-1}$ of luminosity does not
have sensitivity to the MUM model. This is due to the fact that
it has small couplings, and requires a $b$-quark in the initial state. On the
other hand, in the MDM model the $Z^\prime$ couples to other flavours of quark, so
that the production cross-sections are substantially larger. In particular, we find that the larger cross-sections for the MDM model are driven by $bb \to Z' \to \mu^+ \mu^-$ due to the enhanced $g_{bb}$ coupling.
We show the reach
for the MDM model in Fig~\ref{fig:sensHLLHC}. The solid lines show the regions
of $2\sigma$ and $3\sigma$ exclusion.  The region ruled out by neutrino trident
production is shown in green, and in red the region ruled out by  the $B_s$
mixing constraint~\cite{Arnan:2016cpy}. While the LHC would have sensitivity
in this region, we have not set limits there since it is already ruled
out. The more recent determination~\cite{DiLuzio:2017fdq} is shown as the
blue-dashed line. Finally, the grey dashed lines show the values of the
relative decay width $\Gamma_{Z^\prime}/M_{Z^\prime}$. The reach extends out
to 5~TeV, and couplings $g_{\mu\mu}\approx 2$, corresponding to a width of
10\%. These results are broadly in agreement with the previous projections
in~\cite{Allanach:2017bta}, but not as optimistic due to our taking width and
detector effects into account. 

The 27~TeV HE-LHC proposal has  sensitivity to both the MUM and MDM
models, as shown in Fig.~\ref{fig:sensHELHC}. The HE-LHC could probe masses of
up to 12~TeV and widths of up to 60\% in the MDM model, and masses up to 6~TeV for the MUM model for narrow widths.
The exclusion contours in the Fig~\ref{fig:sensHELHC} have quite
different shapes, which stems from the different mixings and couplings in each
model, the requirement that the couplings can explain the flavour anomalies
and the  different widths over the parameter spaces (as in
Fig~.\ref{fig:gbs}). The lack of sensitivity at low masses is due to larger backgrounds that reduce the sensitivity to large widths.

\begin{figure}
  \begin{center}
\unitlength=15cm
\begin{picture}(1.0,0.5)(0,0)
\put(0,0){\includegraphics[width=0.5\textwidth]{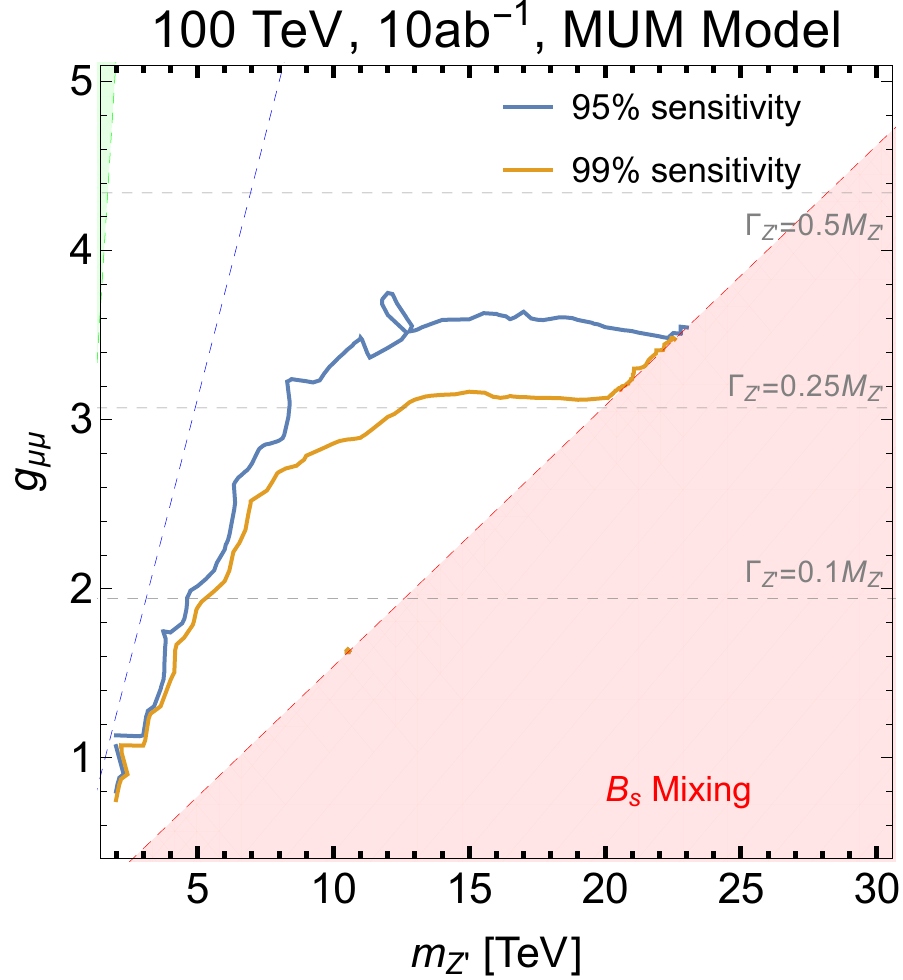}}
\put(0.5,0){\includegraphics[width=0.62\textwidth]{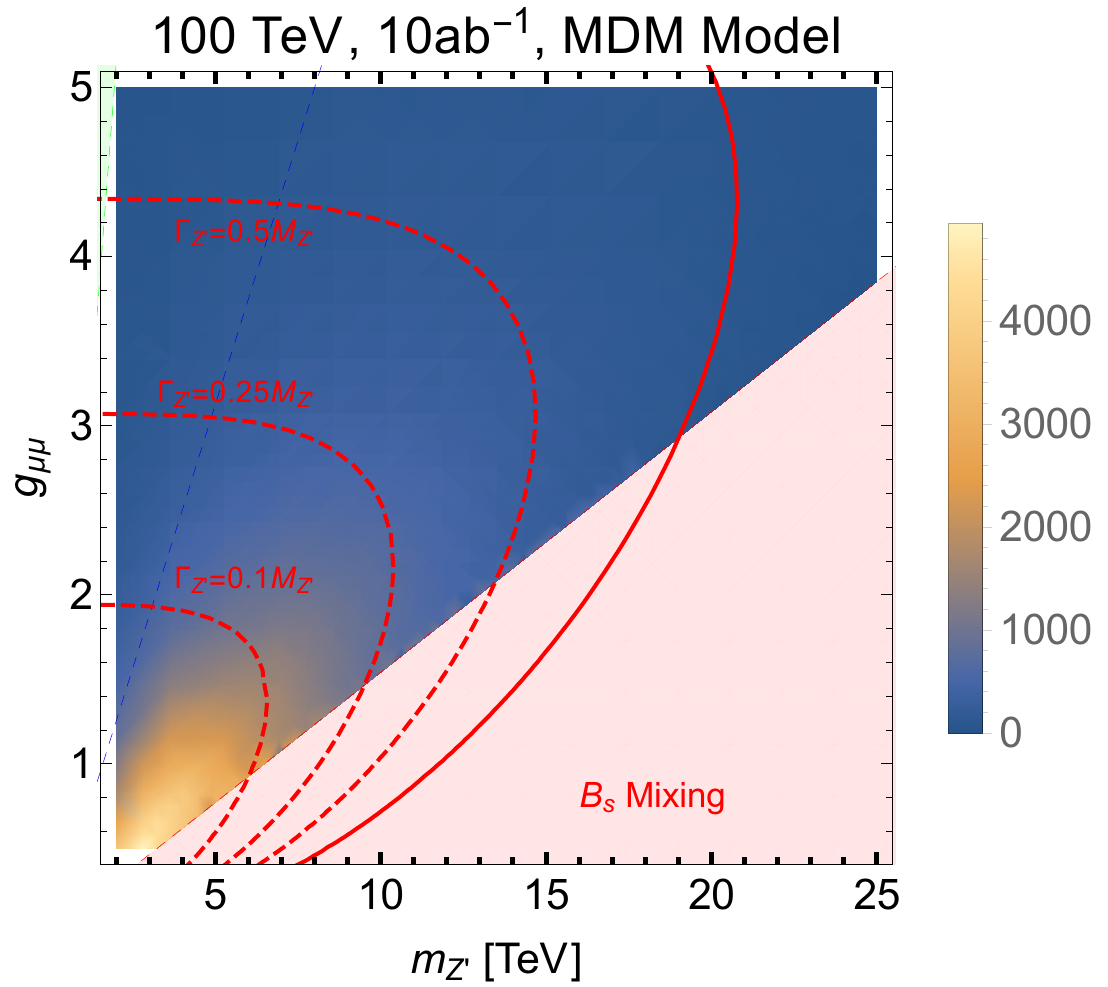}}
\put(0.1,0.075){\sf \footnotesize Allanach, Corbett, Dolan, You, 2018}
\put(0.65,0.075){\sf \footnotesize Allanach, Corbett, Dolan, You, 2018}
\end{picture}
  \end{center}
\caption{\label{fig:sensFCC} Predicted sensitivity for
  the 100~TeV FCC with 10/ab integrated luminosity in the
  $M_{Z^\prime}$--$g_{\mu\mu}$ plane for the MUM model (left). 
  Each point in the plane fits the neutral current
  $B-$anomalies, since Eq.~\ref{constraint} with $x=1.00$ has been enforced
  meaning that $g_{bs}   \propto M_{Z^\prime}^2/g_{\mu\mu}$.
  The solid lines in the left-hand panel
  show the regions of   95$\%$ CL and 99$\%$ CL sensitivity being {\em
  below}\/ each contour. The 
  region ruled out by the $B_s$ mixing constraint~\cite{Arnan:2016cpy} is
  shown in red while the region derived
  from Ref.~\cite{DiLuzio:2017fdq} would 
  be below the blue-dashed line.  
  The grey dashed lines show the values of the relative decay
  width $\Gamma_{Z^\prime}/M_{Z^\prime}$. The MDM model's significance $S$ is colour-coded with the legend on the right; its sensitivity is $S>10\sigma$ everywhere on the plane (right).}  
\end{figure}

Finally, Fig.~\ref{fig:sensFCC} shows the predicted sensitivity in the
$M_{Z^\prime}$--$g_{\mu\mu}$ plane for the FCC with $10\text{ab}^{-1}$
integrated luminosity. We show the MUM model on the left in the range
$M_{Z^\prime}\leq 30$~TeV and $g_{\mu\mu}\leq 5$~TeV.  The FCC has sensitivity to the MUM model in
parameter space not currently ruled out for $M_{Z^\prime}$ up to 23~TeV and for
widths up to 35\%, corresponding to $g_{\mu\mu}\sim 3.5$.
{\em All}\/ of the MDM parameter plane in the right-handed panel of
Fig.~\ref{fig:gbs} is above 10$\sigma$ significance. However, for the
region of large $M_{Z^\prime} \gtrsim 18$ TeV {\em and}\/ $g_{\mu\mu} \gtrsim
3$, $\Gamma_{Z^\prime}\geq M_{Z^\prime}$ and so perturbation
theory is no longer valid.

\section{Conclusion}
\label{sec:conc}

\begin{figure}
  \begin{center}
\unitlength=15cm
\begin{picture}(1.0,0.5)(0,0)
\put(0,0){\includegraphics[width=0.5\textwidth]{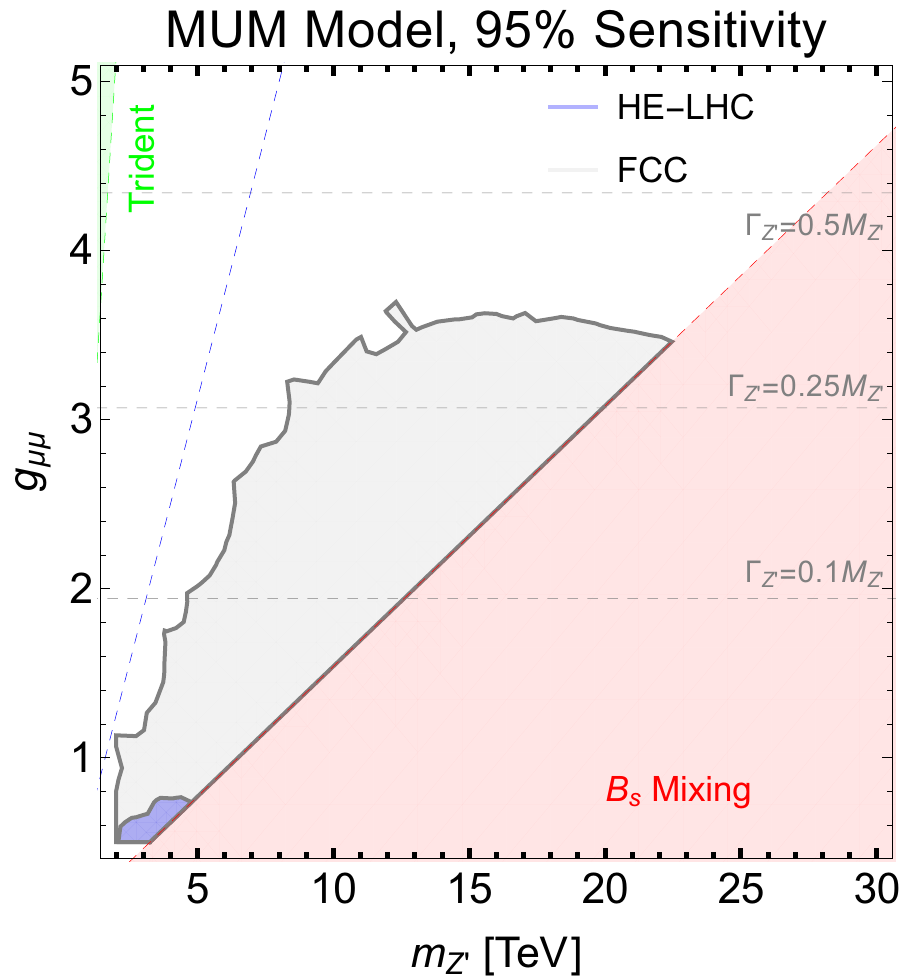}}
\put(0.5,0){\includegraphics[width=0.5\textwidth]{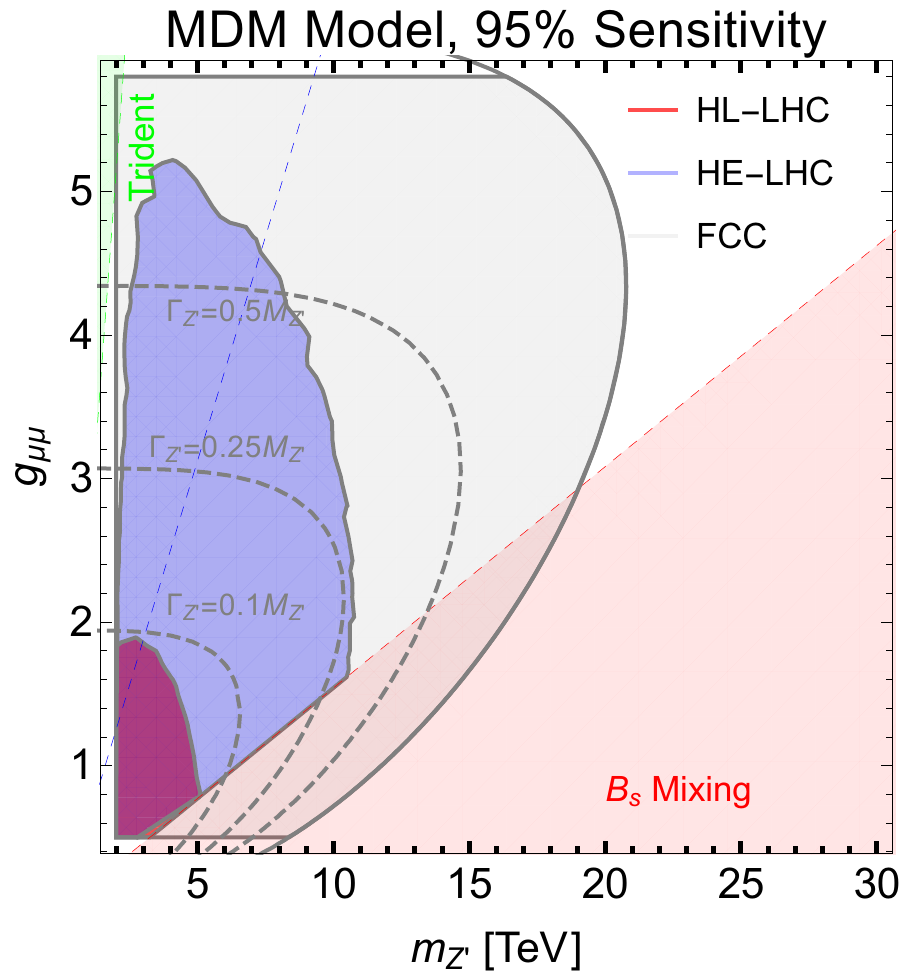}}
\put(0.15,0.075){\sf \footnotesize Allanach, Corbett, Dolan, You, 2018}
\put(0.65,0.075){\sf \footnotesize Allanach, Corbett, Dolan, You, 2018}
\end{picture}
\end{center}
\caption{\label{fig:summary} Summary of MUM model and MDM model 95$\%$
  sensitivities for  
future hadron colliders: 
  the 100~TeV FCC with 10/ab integrated luminosity, 27~TeV HE-LHC with 15/ab
  and the 14~TeV LHC with 3/ab integrated luminosity.
  Each point in the plane fits the neutral current
  $B-$anomalies, since Eq.~\ref{constraint} with $x=1.00$ has been enforced
  meaning that $g_{bs}   \propto M_{Z^\prime}^2/g_{\mu\mu}$.
The solid lines show the regions of 
 95$\%$ CL and 99$\%$ CL sensitivity being {\em
  below}\/ each contour. The 
  region ruled out by the $B_s$ mixing constraint~\cite{Arnan:2016cpy} is
  shown in red while the region derived
  from Ref.~\cite{DiLuzio:2017fdq} would 
  be below the blue-dashed line.  
  The grey dashed lines show the values of the relative decay
  width $\Gamma_{Z^\prime}/M_{Z^\prime}$. The FCC reach is shown in grey and extends throughout the whole perturbative
region where $\Gamma_{Z^\prime}/M_{Z^\prime}<1$ in the MDM model.
} 
\label{fig:comb}
\end{figure}
The $R_{K^{(\ast)}}$ flavour anomalies (discrepancies between SM predictions
and experimental 
measurements in certain $B-$meson decays) are of considerable current interest
and, at face 
value, require the 
existence of physics beyond the SM\@. One possibility involves the existence of a
$Z^{\prime}$, a new heavy vector-boson with flavour dependent couplings. If
the anomalies are confirmed it 
would be desirable to directly produce and identify whatever new particles are
responsible at a current or future collider. In this work we have estimated the
sensitivities of the HL-LHC, HE-LHC and FCC proposals to new, flavour-violating
$Z^\prime$s capable of explaining the flavour anomalies. 

The neutral current $B-$anomalies may require large
$Z^\prime$ couplings depending on $M_{Z^\prime}$, and hence involve resonances
with large decay widths: 
fat, flavourful $Z^{\prime}$s. These widths are larger than what is usually
considered in current LHC searches. 

We have developed $SU(2)_L$ respecting simplified models which include the
couplings necessary to  
explain $R_K$ and $R_{K^{(\ast)}}$ (and related) measurements. 
We
pursued two models: the MUM and MDM scenarios. These
differ in whether CKM mixing occurs in the up-quark or the down-quark sector. Our
projections improve upon previous work~\cite{Allanach:2017bta} by including a dynamical width for the
resonance and by modelling detector acceptance and efficiency effects. 
Although we have presented our results strictly in terms of the MUM and MDM models,
any future dedicated studies should bear in mind that the width could
be larger than predicted in the model in question by the $Z^\prime$ having
more couplings than just those in Eq.~\ref{secSU2}. Therefore, the width could
be kept as an additional free parameter in any such studies. 
Generally, the MDM model has far more sensitivity than the MUM model. Although
the additional valence 
quark 
couplings are CKM suppressed in the MDM model as compared to $g_{bs}$, the
coupling to $b \bar b$ is CKM {\em   enhanced}\/: a factor $1/|V_{ts}| \approx
25$ larger than $g_{bs}$. $Z^\prime$ production is then dominantly via $b \bar b
\rightarrow Z^\prime$, so it would be important to pin down the
$b-$quark parton distribution functions as well as possible in order to reduce
the theoretical uncertainty in the production cross-section. 

Our main results are combined in Fig.~\ref{fig:summary}, which shows the
projected reach of our chosen colliders 
in the $M_{Z^\prime}$--$g_{\mu\mu}$ plane. In order to achieve 95$\%$ CL
sensitivity 
requires going beyond the HL-LHC to a higher energy machine for the MUM
model.  Higher
energy colliders also 
have substantially increased mass reach for 
these resonances: up to 23~TeV for a resonance with 35\% width in the MUM
scenario. We note the importance of the accurate estimation of important
non-perturbative parameters used as inputs to $B_s - \overline{B_s}$
constraints: the default bound is shown by the pink triangular region to the
lower right hand side of the plot, but one recent determination would move the
bound instead to be below the blue dashed line, removing all viable parameter
space where one has 95$\%$ sensitivity to
the MUM model, for instance.

We await confirmation~\cite{Albrecht:2017odf} of 
$R_{K^{(\ast)}}$ flavour anomalies by analyses of LHCb Run II data and
an independent check from Belle II~\cite{Kou:2018nap}. 
In the event of such a confirmation, our work makes the conclusion of
Ref.~\cite{Allanach:2017bta} more robust, and
extends it to 
the large width $Z^\prime$ case: 
the neutral current flavour anomalies and sensitivity to $Z^\prime$ particles
provide another good motivation to  
the already strong case for future high-energy hadron colliders.

\section*{Acknowledgements}
We thank other members of the Cambridge SUSY Working Group and A Greljo for
helpful advice and comments and M Mangano for suggesting the project. 
This work has been partially supported by STFC consolidated grant
ST/P000681/1. TY is supported by a Branco Weiss Society in Science Fellowship
and a Research Fellowship from 
Gonville and Caius College, Cambridge. TC and MJD are supported by the Australian Research Council.

\appendix
\section{Field Definitions \label{sec:def}}
We use the following field definitions in terms of representations of
$SU(3)\times SU(2)_L \times U(1)_Y$: 
\begin{eqnarray}
{\bf Q_L'}=(3,\ 2,\ +1/6),\qquad& {\bf L_L'}=(1,\ 2,\ -1/2),\qquad& {\bf
                                                                    e_R'}=(1,\
                                                                    1,\ 
-1) \nonumber \\ {\bf d_R'}=(3,\ 1,\ 
-1/3),\qquad & {\bf u_R'}=(3,\ 1,\ +2/3),\qquad & \phi=(1,\ 2,\ +1/2)
                                                  \nonumber \\ 
&Z^\prime=(1\ ,1\ ,0).\qquad&
 \end{eqnarray}

\bibliographystyle{JHEP-2}
\bibliography{wzp}

\end{document}